\documentclass{ptephy}

\usepackage{amsmath}
\usepackage{bm}
\usepackage{mathrsfs}

\begin{document}

\title{Constraint on Pulsar Wind Properties from Induced Compton Scattering off Radio Pulses}

\author{\name{\fname{Shuta} \midname{J.} \surname{Tanaka}}{1} and \name{\fname{Fumio} \surname{Takahara}}{2}} 

\address{
\affil{1}{Institute for Cosmic Ray Research, University of Tokyo, 5-1-5 Kashiwa-no-ha, Kashiwa City, Chiba, 277-8582, Japan}
\affil{2}{Department of Earth and Space Science, Graduate School of Science, Osaka University, 1-1 Machikaneyama-cho, Toyonaka, Osaka 560-0043, Japan}
\email{sjtanaka@icrr.u-tokyo.ac.jp}
}

\begin{abstract}%
Pulsar winds have longstanding problems in energy conversion and pair cascade processes which determine the magnetization $\sigma$, the pair multiplicity $\kappa$ and the bulk Lorentz factor $\gamma$ of the wind.
We study induced Compton scattering by a relativistically moving cold plasma to constrain wind properties by imposing that radio pulses from the pulsar itself are not scattered by the wind as was first studied by Wilson \& Rees.
We find that relativistic effects cause a significant increase or decrease of the scattering coefficient depending on scattering geometry.
Applying to the Crab, we consider uncertainties of an inclination angle of the wind velocity with respect to the radio beam $\theta_{\rm pl}$ and the emission region size $r_{\rm e}$ which determines an opening angle of the radio beam.
We obtain the lower limit $\gamma\gtrsim10^{1.7}r^{1/2}_{\rm e,3}\theta^{-1}_{\rm pl}(1+\sigma)^{-1/4}$ ($r_{\rm e}=10^3r_{\rm e,3}$ cm) at the light cylinder $r_{\rm LC}$ for an inclined wind $\theta_{\rm pl}>10^{-2.7}$.
For an aligned wind $\theta_{\rm pl}<10^{-2.7}$, we require $\gamma>10^{2.7}$ at $r_{\rm LC}$ and an additional constraint $\gamma>10^{3.4}r^{1/5}_{\rm e,3}(1+\sigma)^{-1/10}$ at the characteristic scattering radius $r_{\rm c}=10^{9.6}r^{2/5}_{\rm e,3}$ cm within which the `lack of time' effect prevents scattering.
Considering the lower limit $\kappa\gtrsim10^{6.6}$ suggested by recent studies of the Crab Nebula, for $r_{\rm e}=10^3$ cm, we obtain the most optimistic constraint $10^{1.7}\lesssim\gamma\lesssim10^{3.9}$ and $10^{6.6}\lesssim\kappa\lesssim10^{8.8}$ which are independent of $r$ when $\theta_{\rm pl}\sim1$ and $1+\sigma\sim1$ at $r_{\rm LC}$.
\end{abstract}

\subjectindex{xxxx, xxx}

\maketitle

\section{Introduction}

Pulsar magnetospheres create pulsar winds through pair creation and particle acceleration \cite{gj69}.
Because pulsar winds are radiatively inefficient, it is difficult to constrain their properties.
However, their properties are inferred from observations of surrounding pulsar wind nebula (PWN) and pulsed emissions of the pulsar itself.
Interestingly, a secular increase of their pulse period tells us their total energy output $L_{\rm spin}$.
Because most of the spin-down power is converted into the pulsar wind, $L_{\rm spin}$ constrains its properties as (see also Equation (\ref{eq:Density}))
\begin{equation}\label{eq:SpinDownPower}
\kappa \gamma (1 + \sigma) = 1.4 \times 10^{10} \left( \frac{L_{\rm spin}}{10^{38}~\rm erg~s^{-1}} \right)^{\frac{1}{2}},
\end{equation}
where $\kappa$ is the pair multiplicity ($e^{\pm}$ number flux normalized by the Goldreich-Julian number flux $\dot{N}_{\rm GJ}$), $\gamma$ is the bulk Lorentz factor, and $\sigma$ is the magnetization parameter (the ratio of the Poynting to the kinetic energy fluxes) of the pulsar wind, respectively.
We used $\dot{N}_{\rm GJ} \equiv 2 \pi r^2_{\rm pc} c n_{\rm GJ}(r_{\rm pc}) = \sqrt{6 c L_{\rm spin}} / e$, where $r_{\rm pc}$ is the polar cap radius, $n_{\rm GJ}(r_{\rm pc})$ is the Goldreich-Julian density at an magnetic pole and the numerical factor two comes from the north and south magnetic poles.
Pair cascade models within the magnetosphere of the Crab pulsar ($L_{\rm spin} = 4.6 \times 10^{38}~\rm erg~s^{-1}$) predict $\kappa \sim 10^4$ with $\sigma \sim 10^4$ and $\gamma \sim 10^2$ in the vicinity of the light cylinder $r_{\rm LC}$ \cite[e.g.,][]{dh82, ha01, h06}.
On the other hand, magnetohydrodynamic (MHD) models of the Crab Nebula reproduce its non-thermal emission from optical to $\gamma$-ray with $\kappa \sim 10^4$, $\sigma \sim 10^{-3} - 10^{-2}$ and $\gamma \sim 10^6$ \cite{kc84a, kc84b, det06, vet08}.
Although $\kappa \sim 10^4$ in both models is consistent with particle number conservation, $\sigma$ (and also $\gamma$) differs by many orders of magnitude, which is called the `$\sigma$-problem' \cite[c.f.,][]{ket09}.

It is noted that there is an additional problem of the pulsar wind properties \cite[c.f.,][]{tt10, a12}.
Because the MHD models of the Crab Nebula do not explicitly account for the origin of radio emitting particles, they may underestimate the pair multiplicity.
Recent studies of spectral evolution of PWNe showed $\kappa > 10^6$ for the Crab Nebula and $\kappa > 10^5$ for other PWNe \cite[e.g,][]{tt10, tt11, tt13, bet11}.
Although the origin of the low energy particles that are responsible for the radio emission of PWNe is still an open problem, they originate most likely from the pulsar because of the continuity of the broadband spectrum and because of the radio structures apparently originating from the pulsar \cite{aet00, bet01, bet04}.
Thus there arises another problem on $\kappa$ besides the $\sigma$-problem, while only the combination of $\kappa \gamma (1 + \sigma)$ in Equation (\ref{eq:SpinDownPower}) is firm. 

In view of the $\sigma$- and $\kappa$-problems, it is interesting to consider other independent constraints on the physical conditions of pulsar winds.
Wilson \& Rees (1978, hereafter WR78) \cite{wr78} considered induced Compton scattering off radio pulses by a pulsar wind. 
So far, it is thought that we have not observed a signature of scattering in radio spectra of pulsars, although we do not fully understand how scattering changes the radio spectrum (e.g., scattering by a non-relativistic plasma was studied by \cite{zs72, cet93}).
Observations suggest that the optical depth to induced Compton scattering is less than unity, and the radio spectrum is not changed.
Based on this consideration, WR78 obtained the lower limit of the bulk Lorentz factor of the Crab pulsar wind $\gamma > 10^4$ at $10^3 r_{\rm LC} \sim 10^{11}$ cm away from the pulsar.
Substituting Equation (\ref{eq:SpinDownPower}), only for $(1 + \sigma) \sim 1$ at $10^3 r_{\rm LC}$, their conclusion is marginally consistent with the conclusion of $\kappa \gtrsim 10^{6.6} \equiv \kappa_{\rm PWN}$ obtained from the study of the Crab Nebula spectrum by Tanaka \& Takahara (2010, 2011) \cite{tt10, tt11}.

Induced Compton scattering process has been studied for the application to high brightness temperature radio sources, such as the pulsars \cite[e.g.,][]{wr78, lp96, p08a, p08b}, active galactic nuclei \cite[e.g.,][]{s71, cet93, sc96} and other sources \cite[e.g.,][]{zet72, bet90, l08}.
Induced Compton scattering is about a factor of $\theta^4_{\rm bm} k_{\rm B} T_{\rm b}(\nu) / m_{\rm e} c^2$ times effective compared with spontaneous one in the rest frame of the plasma, where $\theta_{\rm bm}~(< 1)$ and $T_{\rm b}$ are a half-opening angle and a brightness temperature of a radio beam, respectively (see Equation (\ref{eq:NonRelativisticLimit})).
Note that the value of $k_{\rm B} T_{\rm b}(\nu) / m_{\rm e} c^2$ can be larger than $10^{15}$ for the Crab pulsar (see Equation (\ref{eq:BrightnessTemperature})).
However, for scattering by relativistically moving electrons, the scattering coefficient is modified by relativistic effects and, as we will see below, either an increase or a decrease is possible depending on situations considered, e.g., the velocity ${\bm u} = \gamma {\bm \beta}$ of the electrons and an inclination between an electron motion ${\bm u}$ and a radio beam ${\bm k}$, where ${\bm k}$ is the wavenumber vector.

In this paper, we reconsider induced Compton scattering by a relativistically moving plasma and reevaluate a lower limit of the bulk Lorentz factor.
Despite strong dependence on scattering geometry, WR78 considered a specific scattering geometry where the pulsar wind is completely aligned with respect to the radio pulse beam and where $\theta_{\rm bm}$ of the radio beam is the widest value inferred from the observations.
We consider rather general geometries of the system, such as the direction of the wind being inclined with respect to the radio pulse beam.
Even if the direction of pulsed radio emission is almost radial, the pulsar wind is likely to have a significant toroidal velocity just outside $r_{\rm LC}$, or its motion in the meridional plane is not strictly radial.
As already noted by WR78, the scattering coefficient may be significantly reduced if the pulsar wind inclines with respect to the radio beam.
For $\theta_{\rm bm}$, the scattering coefficient is reduced when the radio beam is narrow in the rest frame of the plasma.
If this is the case, the lower limit of the bulk Lorentz factor of the pulsar wind may be reduced so as to be consistent with recent studies of the Crab Nebula spectrum.

While we focus on geometrical effects in this paper, we ignore effects of the magnetic field and background photons following WR78.
The magnetic field effect may be important when the frequency of the photon at the plasma rest frame $\nu'$ is smaller than the electron cyclotron frequency $\nu_{\rm ce}$ \cite[e.g.,][]{bs76, lp96}.
For the Crab pulsar, although the magnetic field in the observer frame is about $B_{\rm obs} \sim 10^6 \rm G$ at the light cylinder ($\nu_{\rm ce} = 5.8 \times 10^{12} {\rm Hz}$ for the magnetic field of $B' = 10^6$ G in the plasma rest frame), $\nu_{\rm ce}$ strongly depends on the magnetic field configuration and a direction of plasma motion in the observer frame.
For example, if ${\bm B}_{\rm obs} \perp {\bm u}$, we find $B' = B_{\rm obs}/\gamma$ and $\nu' = \nu / \delta_{\rm D}$ where $\delta_{\rm D}$ is the Doppler factor.
Basically, the magnetic field effect reduces the scattering cross section, i.e., smaller $\gamma$ would be allowed.
For the effect of background photons, Lyubarsky \& Petrova (1996) \cite{lp96} discussed that scattering off the background photons induced by the beam photons may be important.
They discussed that the occupation number of the background photons increases exponentially, i.e., the beam photons may decrease accordingly, when the scattering optical depth to the background photons well exceeds unity, say $10^2$.
In this paper, we ignore background photons ($\theta_{\rm bm} < \theta \le \pi$) assuming that the occupation number of the beam photons is much larger than that of the background photons.
If scattering off the background photons is efficient, scattering would be more efficient and larger $\gamma$ would be required.
These processes will be discussed in a separated paper.

In Section \ref{sec:InducedComptonScattering}, we describe the scattering coefficient of induced Compton scattering by a relativistically moving plasma in a general geometry.
We also show simple analytic forms of the scattering coefficient in some specific geometries.
In general geometry, the scattering coefficient is written in an integral form and is obtained numerically in Appendix \ref{app:NumericalIntegration}.
In Section \ref{sec:Crab}, we consider induced Compton scattering at pulsar wind regions, specifically applying to the Crab pulsar.
We show the resultant lower limits of $\gamma$ and also discuss the corresponding upper limits of the pair multiplicity $\kappa$.
We summarize the present results in Section \ref{sec:Summary}.

\section{INDUCED COMPTON SCATTERING OFF A PHOTON BEAM}\label{sec:InducedComptonScattering}

Here, we express the scattering coefficient at a certain position ${\bm x}$ and see that the scattering coefficient strongly depends on geometry of scattering.
The kinetic equation for a photon occupation number $n({\bm x}, {\bm k}, t)$ is expressed as \cite[e.g.,][]{hm95, lp96}
\begin{eqnarray}\label{eq:Boltzmann}
	\left( \frac{\partial}{\partial (c t)} + {\bm \Omega} \cdot {\bm \nabla} \right) n({\bm k}) 
	& = & 
	\int d^3 {\bm p} f({\bm p}) \int \frac{d^3 {\bm k}_1}{k^2_1} \frac{d \sigma}{d \Omega}({\bm p}, {\bm k}, {\bm k_1}) \left[ n({\bm k}_1) (1 + n({\bm k})) \left(\frac{k_1}{k} \right)^2 \right. \nonumber \\
	& & \left. \delta(k - g({\bm p}, {\bm k_1})) - n({\bm k})(1 + n({\bm k_1})) \delta(k_1 - g({\bm p}, {\bm k})) \right], 
\end{eqnarray}
where ${\bm \Omega} = {\bm k} / k$, $f({\bm p})$ is the distribution function of plasma and $d \sigma/ d \Omega$ is the differential scattering cross section, respectively.
Note that when the electron is initially at rest, the recoil $g$ is expressed as $g(k, \xi) = k / (1 + k \lambda_{\rm e} (1 - \cos \xi))$, where $\lambda_{\rm e} = \hbar / m_{\rm e} c$ represents the Compton wavelength for an electron and $\xi$ is the angle between incident and scattered photons.
We omit arguments ${\bm x}$ and $t$ in Equation (\ref{eq:Boltzmann}) and in this section.
The terms $1 + n$ represent spontaneous and induced scattering terms, and we only consider the induced process below, assuming $n \gg 1$.

\subsection{Scattering Coefficient}\label{sec:ScatteringCoeffiecnt}

The scattering coefficient of induced Compton scattering is the right-hand side of Equation (\ref{eq:Boltzmann}) divided by $n({\bm k})$ \cite[e.g.,][]{w82}.
Equation (\ref{eq:Boltzmann}) is simplified by following three approximations.
(I) Plasma is cold, and moves with the velocity ${\bm u} = \gamma {\bm \beta}$ (the bulk Lorentz factor $\gamma = (1 - \beta^2)^{-1/2}$).
(II) The magnetic field is weak enough to satisfy the condition $\nu_{\rm ce} < \nu'$, where $\nu_{\rm ce}$ and $\nu'$ are the electron cyclotron frequency and the frequency of an incident photon in the plasma rest frame, respectively \cite[e.g.,][]{lp96}.
(III) Photons are in the Thomson regime, i.e., $k \lambda_{\rm e} \ll 1$ \cite[c.f.,][]{hm95}.
The condition (III) is a good approximation for scattering off radio photons by plasma of $\gamma \ll 10^{10}$.
In the observer frame, Equation (\ref{eq:Boltzmann}) then becomes, \cite[e.g.,][]{lp96, wr78},
\begin{eqnarray}\label{eq:InducedScattering}
	\left( \frac{\partial}{\partial (c t)} + {\bm \Omega \cdot \nabla} \right) n({\bm k}) 
	& = & 
	n({\bm k}) \frac{3}{8 \pi} \sigma^{}_{\rm T} n_{\rm pl} \int \frac{d {\bm \Omega}_1}{\gamma^3 D^2_1} R({\bm \Omega}, {\bm \Omega}_1, {\bm u}) (1 - \mu) \lambda_{\rm e} \frac{\partial k^2_1 n({\bm k}_1)}{\partial k_1} \Biggl|_{k_1 = \frac{D}{D_1} k},
\end{eqnarray}
where
\begin{eqnarray}
1 - \mu & = & 1 - {\bm \Omega} \cdot {\bm \Omega}_1, \label{eq:Mu} \\
D       & = & 1 - {\bm \beta}  \cdot {\bm \Omega},   \label{eq:D} \\
D_1     & = & 1 - {\bm \beta}  \cdot {\bm \Omega}_1, \label{eq:D1} \\ 
R({\bm \Omega}, {\bm \Omega}_1, {\bm u}) & = & 1 + \left(1 -  \frac{1 - \mu}{\gamma^2 D D_1} \right)^2, \label{eq:R}
\end{eqnarray}
and $n_{\rm pl}$ is a number density of plasma.
$R({\bm \Omega}, {\bm \Omega}_1, {\bm u})$ is order unity ($1 \le R \le 2$) and $\sigma^{}_{\rm T}$ is the Thomson scattering cross section.
The scattering coefficient contains the integral which depends on the occupation number itself and on scattering geometry at ${\bm x}$, i.e., directions of photons (${\bm \Omega}$ and ${\bm \Omega}_1$) and a velocity of the plasma ${\bm u}$.
While WR78 performed this integral on a specific scattering geometry, we reevaluate it in more general geometries.

\subsection{Geometry}\label{sec:Geometry}

%
\begin{figure}[t]
\centering
\includegraphics[scale=0.4]{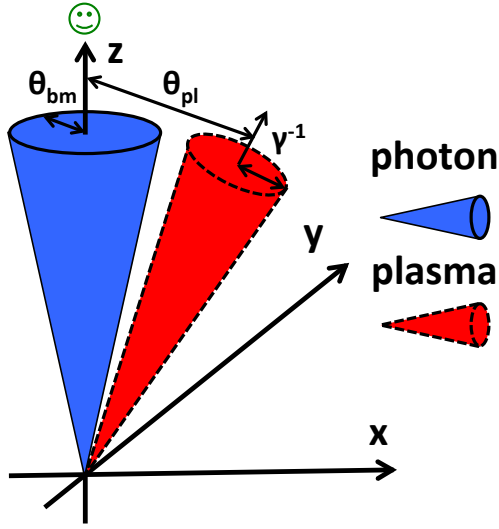}
\caption{ 
	A sketch of scattering geometry.
	The photon beam of a half-opening angle $\theta_{\rm bm}$ is toward the observer on z-axis.
	An inclination angle of the cold plasma is $\theta_{\rm pl}$ ($\phi_{\rm pl} = 0$).
	Although the cold plasma beam should be described as a line, we associate it with the cone of a half-opening angle of $\gamma^{-1}$ in all the sketches for explanatory convenience.  
} 
\label{fig:Geometry}
\end{figure}
Scattering geometry at a certain position ${\bm x}$ in the observer frame is depicted in Figure \ref{fig:Geometry}.
The photon beam with a half-opening angle $\theta_{\rm bm}$ directs to an observer on $z$-axis.
An inclination angle of the plasma velocity is $\theta_{\rm pl}$.
Note that the plasma should be depicted as a line rather than a cone on Figure \ref{fig:Geometry}, i.e., zero opening angle, because we assume that the plasma is cold. 
However, we will see that there is the characteristic angle $\gamma^{-1}$ around the plasma velocity and then we associate the plasma with the cone of its half-opening angle $\gamma^{-1}$ in the figures in this paper.

For the plasma, we express the velocity ${\bm u}$ as
\begin{equation}\label{eq:PlasmaDistribution}
{\bm u} = \gamma \beta (\sin \theta_{\rm pl} {\bm e}_x + \cos \theta_{\rm pl} {\bm e}_z).
\end{equation}
We assume that the occupation number of photons is uniform inside the beam and is expressed as
\begin{eqnarray}\label{eq:PhotonAngularDistribution}
n({\bm k}) & = & n(\nu, {\bm \Omega}) \nonumber \\
	   & = & n(\nu) H(\theta_{\rm bm} - \theta),
\end{eqnarray}
where $H$ is the Heaviside's step function.
The spectrum $n(\nu)$ is assumed to be a broken power-law form
\begin{equation}\label{eq:PhotonFrequencyDistribution}
n(\nu) = n_0 \left( \frac{\nu}{\nu_0} \right)^{p_1} \left[ \frac{1}{2} \left( 1 + \frac{\nu}{\nu_0} \right) \right]^{p_2 - p_1},
\end{equation}
where $p_1$ and $p_2$ are power-law indices of low and high frequency parts and $n_0$ is the occupation number at a break frequency $\nu_0$, respectively. 
Observed pulsar radio spectra correspond to $-7 \lesssim p_2 \lesssim -3$, and we require $p_1 > -3$ for the number density of photons to be finite at $\nu \rightarrow 0$.
For the application in Section \ref{sec:Crab}, we take $p_2 = -5$ and $\nu_0 = 10$ MHz considering the radio observations.
Adopting $p_1 = 3$, the brightness temperature $k_{\rm B} T_{\rm b}(\nu) = h \nu n(\nu)$ to be maximum at $\nu_0$.

We consider scattering off photons toward the observer, i.e., ${\bm \Omega} = {\bm e}_z$.
The scattering coefficient $\chi$ at ${\bm x}$ is expressed as
\begin{eqnarray}\label{eq:ScatteringCoefficient}
	\chi(\nu, {\bm e}_z) 
	& \equiv & 
	- \frac{3}{8 \pi} n_{\rm pl} \sigma^{}_{\rm T} \int^{2 \pi}_0 d \phi_1 \int^{\theta_{\rm bm}}_0 \sin \theta_1 d \theta_1 
	\frac{1 - \mu}{\gamma^3 D^2_1} R({\bm e}_z, {\bm \Omega}_1, {\bm u}) 
	\left( \frac{k_{\rm B} T_{\rm b}(\nu_1)}{m_{\rm e} c^2} \frac{\partial \ln n(\nu_1) \nu^2_1}{\partial \ln \nu_1} \right)_{\nu_1 = \frac{D}{D_1} \nu}.
\end{eqnarray}
As is the conventional definition of the optical depth $d \tau = \chi d l$ for a path $l$ along $z$-axis, we include a minus sign, where the occupation number decreases along the path for a positive value of $\chi$ and vice versa.
The sign of $\chi$ can change with the sign of the function
\begin{equation}\label{eq:Sign}
S(\nu) \equiv \frac{\partial \ln n(\nu) \nu^2}{\partial \ln \nu} \approx \left\{
\begin{array}{ll}
p_1 + 2  & \mbox{ for $\nu \ll \nu_0$,} \\
p_2 + 2  & \mbox{ for $\nu \gg \nu_0$.} \\
\end{array} \right.
\end{equation}
%

\subsection{Analytic Estimates}\label{sec:AnalyticEstimates}

It is convenient to rewrite Equation (\ref{eq:ScatteringCoefficient}) by introducing the normalization 
\begin{equation}\label{eq:Normalization}
\chi_0 \equiv n_{\rm pl} \sigma^{}_{\rm T} \frac{k_{\rm B} T_{\rm b}(\nu_0)}{m_{\rm e} c^2}.
\end{equation}
The scattering coefficient becomes
\begin{eqnarray}\label{eq:DefineIntegral}
\chi(\nu, {\bm e}_z) & = &
- \frac{3 \chi_0}{8 \pi \gamma^3} \int^{2 \pi}_0 \int^{\theta_{\rm bm}}_0 \sin \theta_1 d \theta_1 d \phi_1 
\frac{1 - \mu}{D^2_1} R({\bm e}_z, {\bm \Omega}_1, {\bm u}) \left( \frac{T_{\rm b}(\nu_1)}{T_{\rm b}(\nu_0)} S(\nu_1) \right)_{\nu_1 = \frac{D}{D_1} \nu} \nonumber \\
& \equiv & \chi_0 \gamma^{-3} I(\nu, \theta_{\rm bm}, \theta_{\rm pl}, \gamma),
\end{eqnarray}
where the integral $I(\nu, \theta_{\rm bm}, \theta_{\rm pl}, \gamma)$ represents a geometrical effect.
Note that $\chi$ contains a factor of $\gamma^{-3}$ which is independent of scattering geometries.
The value of $I(\nu, \theta_{\rm bm}, \theta_{\rm pl}, \gamma)$ is obtained numerically in general and can take a wide range of values even for a fixed frequency.
The numerical results of the integral $I(\nu)$ for different parameter sets $(\theta_{\rm bm}, \theta_{\rm pl}, \gamma)$ are described in Appendix \ref{app:NumericalIntegration} and are also shortly summarized in the last paragraph of this section.
Below, we describe simple analytic forms of the integral $I(\nu)$ for some special cases.
They help understanding of dependence on $(\theta_{\rm bm}, \theta_{\rm pl}, \gamma)$ and turn out to be useful for applications in the next section.

We first see the non-relativistic limit	$\beta \ll 1$ ($D, D_1 \sim 1$) where the $\theta_{\rm pl}$-dependence can be neglected.
Considering $\theta_{\rm bm} < 1$, we obtain
\begin{eqnarray}\label{eq:NonRelativisticLimit}
	I_{\rm NR}(\nu)
	\approx
	- \frac{3}{4} \frac{T_{\rm b}(\nu)}{T_{\rm b}(\nu_0)} S(\nu) \int^{\theta_{\rm bm}}_0 \theta^3_1 d \theta_1
	=
	- \frac{3}{16} \theta^4_{\rm bm} \frac{T_{\rm b}(\nu)}{T_{\rm b}(\nu_0)} S(\nu).
\end{eqnarray}
where we use $R({\bm e}_z, {\bm \Omega}_1, {\bm u}) \approx 1 + (1 - \theta^2_1 / 2)^2 \approx 2$.
When the photon beam is narrow ($\theta_{\rm bm} \ll 1$), the scattering coefficient can be small.
This is because the number of photons which stimulate the scattering process decreases with $\theta^2_{\rm bm}$ and another factor $\theta^2_{\rm bm}$ comes from the recoil term $\propto 1 - \mu \approx \theta^2_1 / 2$. 
For typical values of $p_1$ and $p_2$, $|I_{\rm NR}(\nu)|$ (i.e., $|\chi^{}_{\rm NR}(\nu)|$) has a peak and changes sign at $\nu \approx \nu_0$.

\begin{figure}[t]
\begin{minipage}{0.5\hsize}
\begin{center}
\includegraphics[scale=0.6]{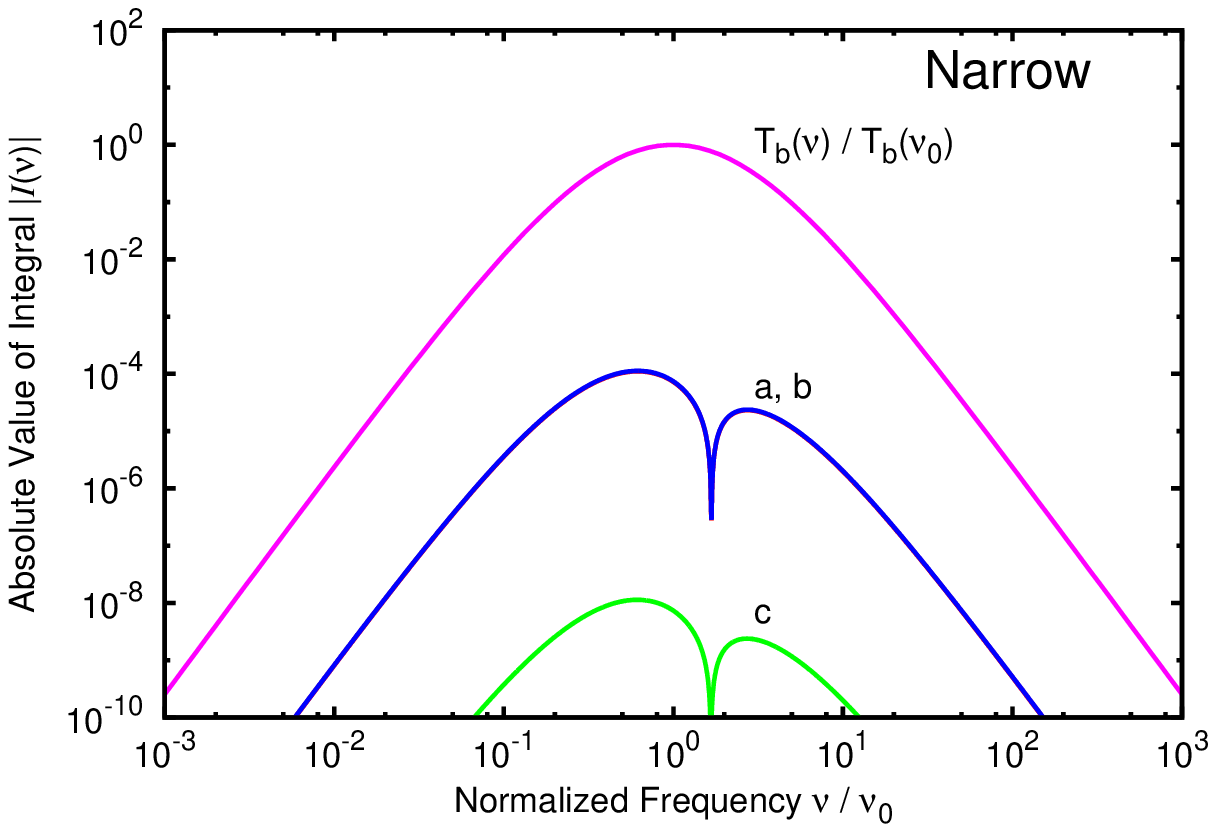}
\end{center}
\end{minipage}
\begin{minipage}{0.5\hsize}
\begin{center}
\includegraphics[scale=0.3]{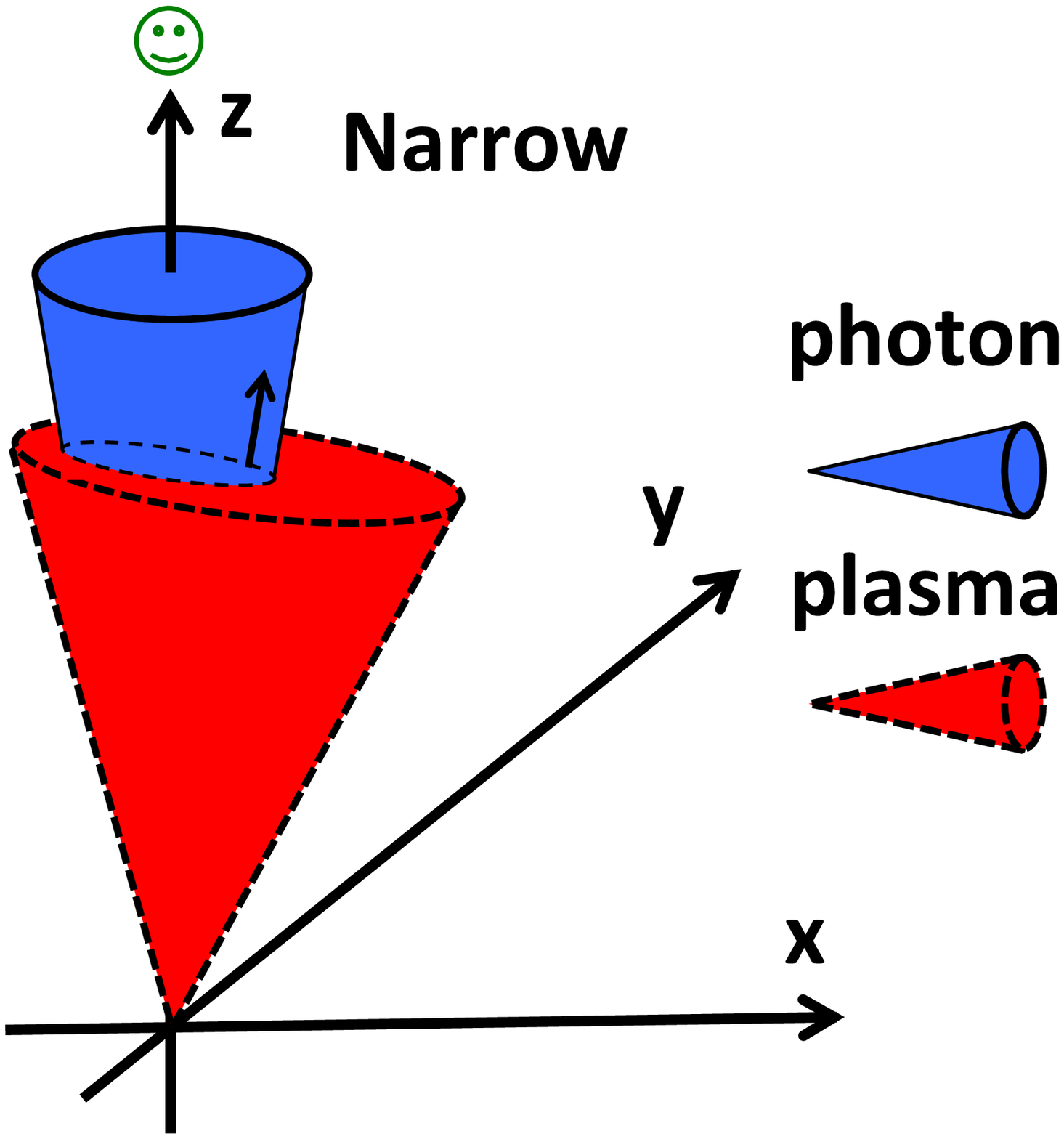}
\end{center}
\end{minipage}
\caption{
Plot of the integral $I(\nu, \theta_{\rm bm}, \theta_{\rm pl}, \gamma)$ in Equation (\ref{eq:DefineIntegral}) (left) and a sketch of scattering geometry (right) in the `Narrow' case ($1 > \Theta^2_{\rm bm} + \Theta^2_{\rm pl}$).
The plot shows absolute values $|I(\nu)|$ versus $\nu$ with $\gamma = 10^2$, $p_1 = 3$ and $p_2 = -5$ (lines a, b and c) together with $T_{\rm b}(\nu) / T_{\rm b}(\nu_0)$ for comparison. 
Each line is for a different value of $(\Theta_{\rm bm}, \Theta_{\rm pl})$: `line a' for $(10^{-1}, 10^{-1})$, `line b' for $(10^{-1}, 10^{-2})$, and `line c' for $(10^{-2}, 10^{-1})$, respectively.
Note that `line a' and `line b' are overlapped since $I(\nu)$ is primarily determined by $\Theta_{\rm bm}$ as is seen in Equation (\ref{eq:NarrowBeam}).
A discontinuity found in each line is the frequency where the sign of $I(\nu)$ changes and the high frequency side has a positive sign, while the low frequency side has a negative sign for all lines.
Note also that, in the right panel, the opening angle of the plasma cone (red in color) represents $\gamma^{-1}$ cone and does not represent the velocity distribution (see Figure \ref{fig:Geometry} and the text).
}
\label{fig:Narrow}
\end{figure}

\begin{figure}[t]
\begin{minipage}{0.5\hsize}
\begin{center}
\includegraphics[scale=0.6]{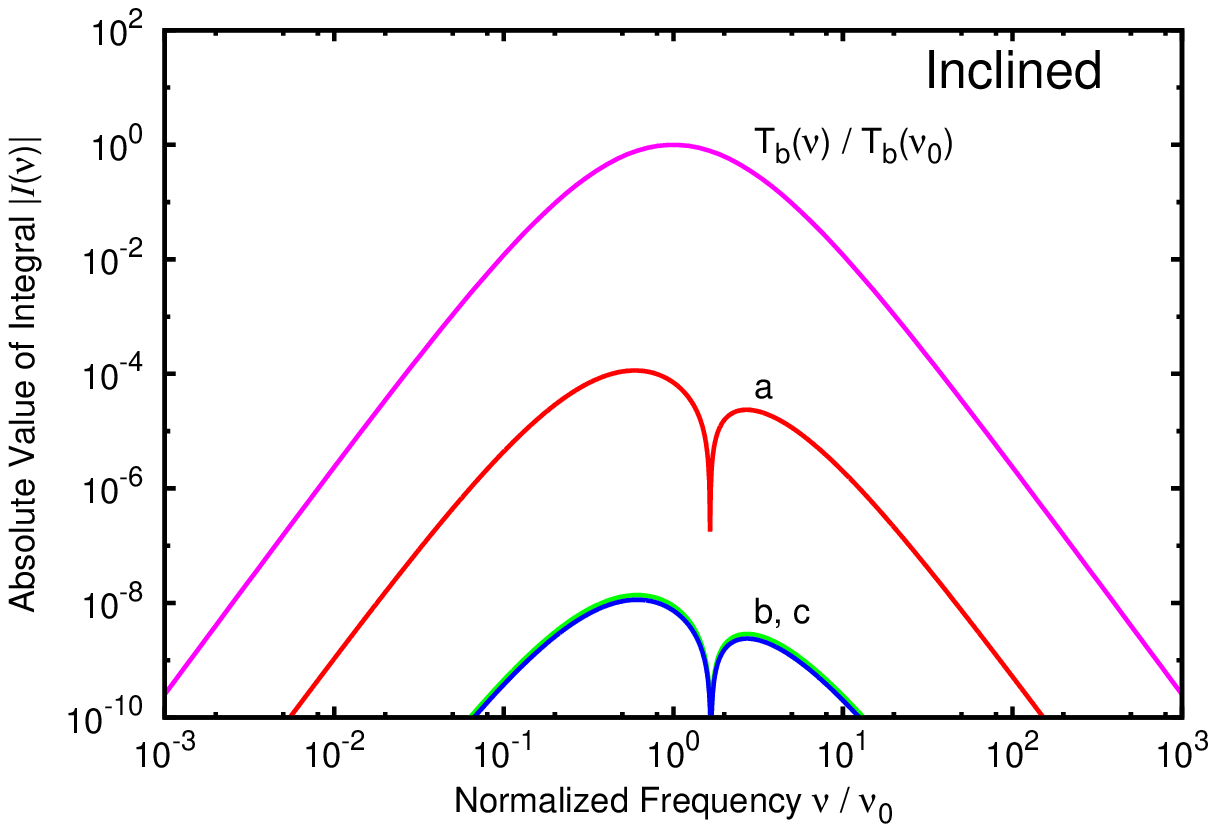}
\end{center}
\end{minipage}
\begin{minipage}{0.5\hsize}
\begin{center}
\includegraphics[scale=0.3]{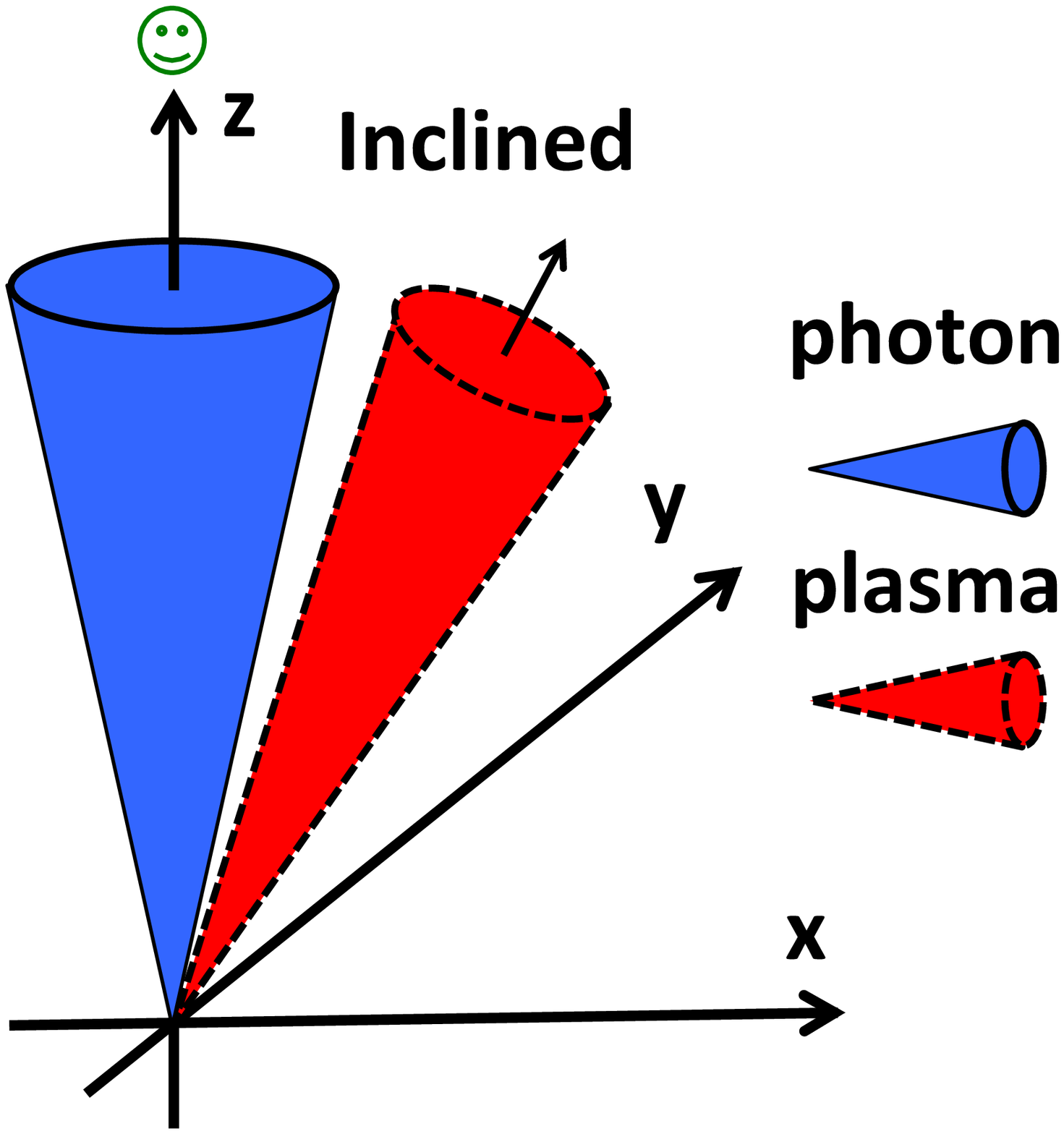}
\end{center}
\end{minipage}
\caption{
Plot of the integral $I(\nu, \theta_{\rm bm}, \theta_{\rm pl}, \gamma)$ in Equation (\ref{eq:DefineIntegral}) (left) and a sketch of scattering geometry (right) in the `Inclined' case ($\Theta^2_{\rm pl} > \Theta^2_{\rm bm} + 1$).
Each line is for a different value of $(\Theta_{\rm bm}, \Theta_{\rm pl})$: `line a' for $(1, 10)$, `line b' for $(1, 10^{2})$, and `line c' for $(10^{-1}, 10)$, respectively and the other parameters are the same as in Figure \ref{fig:Narrow}.
Note that `line b' and `line c' are overlapped since $I(\nu)$ is primarily determined by the ratio $\Theta_{\rm bm} / \Theta_{\rm pl}$ as seen in Equation (\ref{eq:Inclined}).
}
\label{fig:Inclined}
\end{figure}

\begin{figure}[t]
\begin{minipage}{0.5\hsize}
\begin{center}
\includegraphics[scale=0.6]{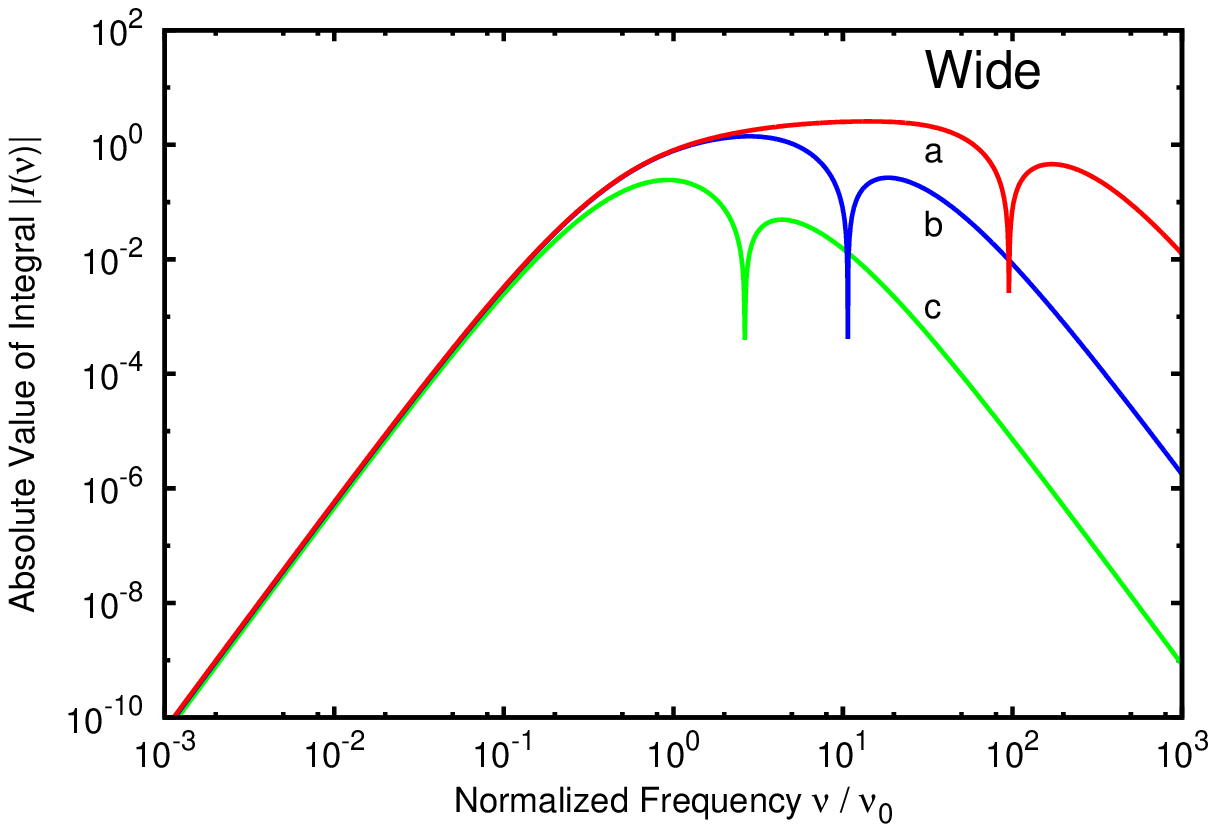}
\end{center}
\end{minipage}
\begin{minipage}{0.5\hsize}
\begin{center}
\includegraphics[scale=0.3]{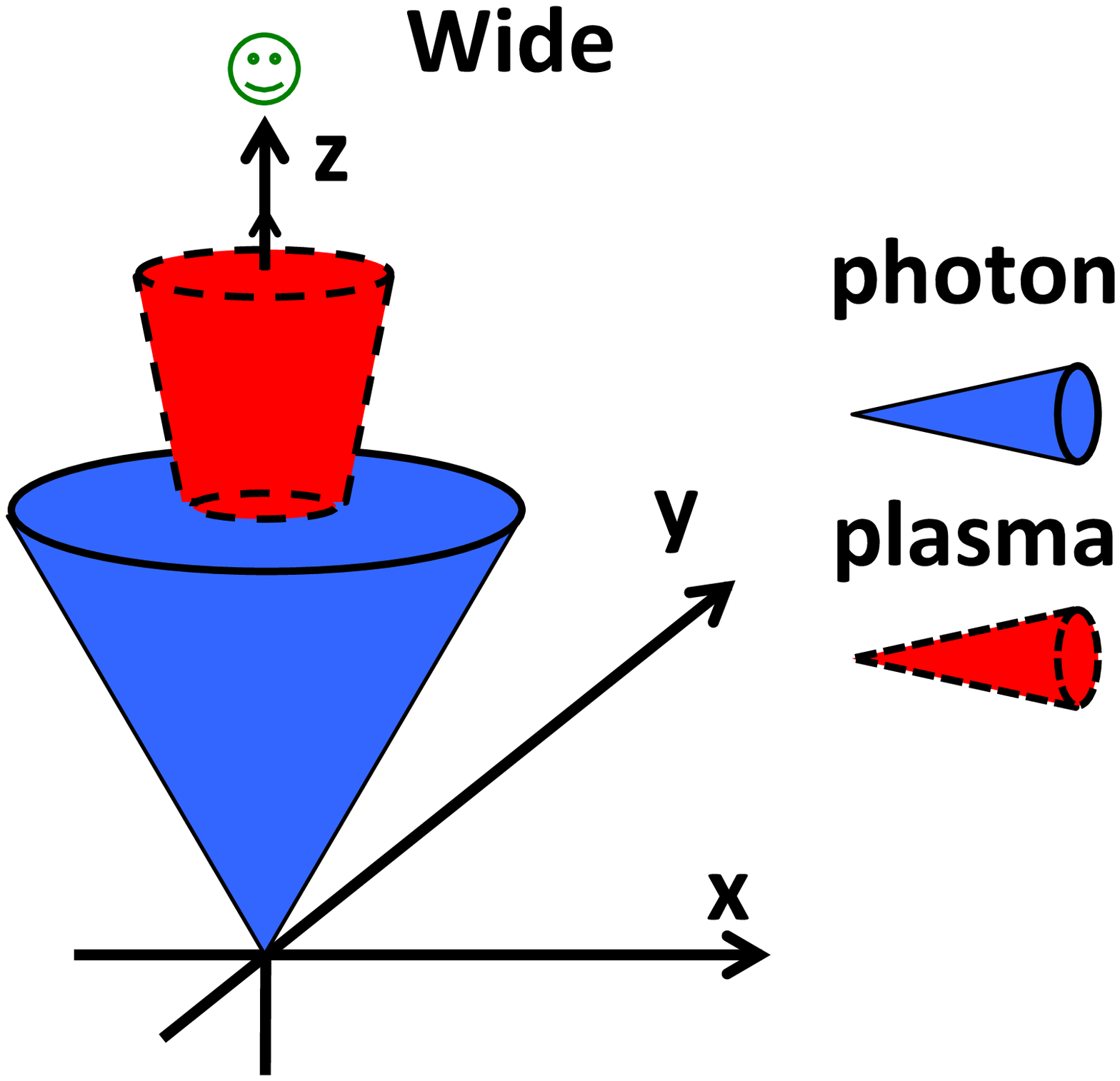}
\end{center}
\end{minipage}
\caption{
Plot of the integral $I(\nu, \theta_{\rm bm}, \theta_{\rm pl}, \gamma)$ in Equation (\ref{eq:DefineIntegral}) (left) and a sketch of scattering geometry (right) in the `Wide' case ($\Theta^2_{\rm bm} > 1 > \Theta^2_{\rm pl} = 0$).
Each line is for a different value of $(\Theta_{\rm bm}, \Theta_{\rm pl})$: where `line a' for $(10, 0)$, `line b' for $(3, 0)$, and `line c' for $(1, 0)$, respectively and the other parameters are the same as shown in Figure \ref{fig:Narrow}.
Note that the discontinuity frequency shifts to higher frequency for larger $\Theta_{\rm bm}$, which matches Equation (\ref{eq:WideBeam1}) well.
}
\label{fig:Wide}
\end{figure}

To see relativistic effects, we expand $\sin \theta$, $\cos \theta$ and $\beta$ to second-order in $\theta_1$, $\theta_{\rm pl}$ and $\gamma^{-1}$, i.e., we concern the situations $0 \le (\theta_{\rm pl},~\theta_{\rm bm}) \lesssim 1$ and $\gamma \gg 1$.
The integrand is composed of following three factors.
(I) The solid angle (and the recoil) factor originates from the solid angle element $d \Omega_1$ and from the recoil term $1 - \mu$, and is expressed as
\begin{equation}\label{eq:SolidAngleFactor}
(1 - \mu) \sin \theta_1 d \phi_1 d \theta_1 \approx \frac{1}{2} \theta^3_1 d \theta_1 d \phi_1.
\end{equation}
This factor already appeared in the non-relativistic case (Equation (\ref{eq:NonRelativisticLimit})).
(II) The aberration factor originates from the Lorentz transformation of a solid angle element from the plasma rest frame to the observer frame, and is expressed as
\begin{equation}\label{eq:AberrationFactor}
\frac{1}{D^2_1} \approx \frac{4 \gamma^4}{(1 + \gamma^2 \psi^2_1)^2},
\end{equation}
where we introduced an angle $\psi_1$ between ${\bm \beta}$ and ${\bm \Omega}_1$, given by the approximation $\psi^2_1 = \theta^2_1 - 2 \theta_1 \theta_{\rm pl} \cos \phi_1 + \theta^2_{\rm pl}$.
(III) The frequency shift factor also originates from the Lorentz transformation of a frequency, and is expressed as 
\begin{equation}\label{eq:FrequencyShiftFactor}
\frac{D}{D_1} \approx \frac{1 + \gamma^2 \theta^2_{\rm pl}}{1 + \gamma^2 \psi^2_1}.
\end{equation}
Analytic forms of the integral $I(\nu)$ presented below are explained by a simple combination of these three factors.
We also show numerical results of the integral $I(\nu)$ for these cases in Figures \ref{fig:Narrow} $-$ \ref{fig:Wide}, where we adopt $p_1 = 3$, $p_2 = -5$ and $\gamma = 10^2$.
Introducing normalized angles $\Theta_{\rm bm} \equiv \gamma \theta_{\rm bm}$ and $\Theta_{\rm pl} \equiv \gamma \theta_{\rm pl}$, it is easy to find that the integral $I(\nu)$ depends on ($\Theta_{\rm bm}, \Theta_{\rm pl})$ rather than separately on $\theta_{\rm bm}$, $\theta_{\rm pl}$ and $\gamma$.

We first consider the case $1 > \Theta^2_{\rm bm} + \Theta^2_{\rm pl}$ where the narrow photon beam and ${\bm \Omega} = {\bm e}_z$ are well inside the $\gamma^{-1}$ cone associated with the plasma as shown in the right panel of Figure \ref{fig:Narrow}.
We call this case `Narrow'.
In this case, we obtain $D^{-2}_1 \approx 4 \gamma^4$ and $D / D_1 \approx 1$, and then the integral $I(\nu)$ is approximated as
\begin{equation}\label{eq:NarrowBeam}
	I_{\rm Narrow}(\nu) 
	\approx
	- \frac{3}{4} \frac{T_{\rm b}(\nu)}{T_{\rm b}(\nu_0)} S(\nu) \int^{\theta_{\rm bm}}_0 (4 \gamma^4) \theta^3_1 d \theta_1
	=
	- \frac{3}{4} \Theta^4_{\rm bm} \frac{T_{\rm b}(\nu)}{T_{\rm b}(\nu_0)} S(\nu),
\end{equation}
where we use $R({\bm e}_z, {\bm \Omega}_1, {\bm u}) \approx 1 + (1 - 2 \Theta^2_1)^2 \approx 2$ ($\Theta_1 \equiv \gamma \theta_1$).
This expression with $\gamma \rightarrow 1$ ($\Theta_{\rm bm} \rightarrow \theta_{\rm bm}$) is almost the same as that of the non-relativistic case (Equation (\ref{eq:NonRelativisticLimit})).
For the `Narrow' case, the aberration factor increases the integral $I(\nu)$ by a factor of $D^{-2}_1 \approx 4 \gamma^4$ compared with $I_{\rm NR}(\nu)$ because the opening angle increases by a factor of $\sim \gamma$ in the plasma rest frame, while the frequency shift is negligible ($D / D_1 \approx 1$).
Note that $\chi^{}_{\rm Narrow}(\nu)$ is a factor of $\gamma$ larger than $\chi^{}_{\rm NR}(\nu)$ accounting for the factor of $\gamma^{-3}$ in Equation (\ref{eq:DefineIntegral}).
In the left panel of Figure \ref{fig:Narrow}, we plot numerical results of absolute values of the integral $I(\nu)$ (Equation (\ref{eq:DefineIntegral})) as a function of $\nu$.
$|I(\nu)|$ has a discontinuity because $S(\nu)$ changes sign at $\nu \sim \nu_0$, where $I(\nu) > 0$ (i.e., $\chi(\nu) > 0$) for $\nu > \nu_0$ and vice versa.

Next case is $\Theta^2_{\rm pl} > \Theta^2_{\rm bm} + 1$ where ${\bm u}$ is inclined with respect to ${\bm \Omega}$ and the associated cones do not overlap with ${\bm \Omega}$ as shown in the right panel of Figure \ref{fig:Inclined}.
We call this case `Inclined'.
The integral $I(\nu)$ also suffers from little frequency shift ($D / D_1 \approx 1$) and the aberration factor is approximated as $D^{-2}_1 \approx 4 \theta^{-4}_{\rm pl}$.
We obtain an approximated form of
\begin{equation}\label{eq:Inclined}
	I_{\rm Inclined}(\nu)
	\approx
	- \frac{3}{4} \frac{T_{\rm b}(\nu)}{T_{\rm b}(\nu_0)} S(\nu) \int^{\theta_{\rm bm}}_0 (4 \theta_{\rm pl}^{-4}) \theta^3_1 d \theta_1
	=
	- \frac{3}{4} \frac{\Theta^4_{\rm bm}}{\Theta^4_{\rm pl}} \frac{T_{\rm b}(\nu)}{T_{\rm b}(\nu_0)} S(\nu),
\end{equation}
where we use $R({\bm e}_z, {\bm \Omega}_1, {\bm u}) \approx 1 + (1 - 2 \Theta^2_1 \Theta^{-4}_{\rm pl})^2 \approx 2$.
In the left panel of Figure \ref{fig:Inclined}, we show numerical results for the `Inclined' case.
The aberration factor decreases the integral $I(\nu)$ by a factor of $\Theta^{-4}_{\rm pl}$ compared with $I_{\rm Narrow}(\nu)$.
Note that $\chi^{}_{\rm Inclined}(\nu)$ can be smaller than $\chi^{}_{\rm NR}(\nu)$, as $\chi^{}_{\rm Inclined}(\nu) / \chi^{}_{\rm NR}(\nu) \sim \gamma^{-3} \theta^{-4}_{\rm pl}$.
For example, we find $\chi^{}_{\rm Inclined}(\nu) \sim \gamma^{-3} \chi^{}_{\rm NR}(\nu)$ for $\theta_{\rm pl} \sim 1$, while $\chi^{}_{\rm Inclined}(\nu) \sim \gamma \chi^{}_{\rm NR}(\nu)$ for $\Theta_{\rm pl} \sim 1$.

The scattering geometry satisfying $\Theta_{\rm bm} > 1 > \Theta_{\rm pl}$ is sketched in the right panel of Figure \ref{fig:Wide} where the $\gamma^{-1}$ cone of plasma contains ${\bm \Omega}$ and is well within the photon beam.
We call this case `Wide'.
Note that although we take $\theta_{\rm pl} = 0$ in Figure \ref{fig:Wide} and in Equation (\ref{eq:WideBeam1}), we will find that the integral $I(\nu)$ behaves in a similar way for $\Theta_{\rm bm} > 1 > \Theta_{\rm pl} \ne 0$ in Appendix \ref{app:NumericalIntegration}.
For $\theta_{\rm pl} = 0$, the frequency shift factor is approximated as $D / D_1 \approx (1 + \Theta^2_1)^{-1} \le 1$.
The aberration factor behave as $D^{-2}_1 \approx 4 \gamma^4 / (1 + \Theta^2_1)^2$ and makes the angular distribution of the photon beam almost isotropic in the plasma rest frame.
Simple analytic form is found for the frequency range $\nu > (1 + \Theta^2_{\rm bm}) \nu_0 \approx \Theta^2_{\rm bm} \nu_0$, where we use the expressions $T_{\rm b}(\nu_1) \approx T_{\rm b}(\nu_0) (\nu/ (1 + \Theta^2_1) \nu_0)^{p_2 + 1}$ and $S(\nu_1) \approx p_2 + 2$.
We obtain an approximated form
\begin{eqnarray}\label{eq:WideBeam1}
	I_{\rm Wide}(\nu > \Theta^2_{\rm bm} \nu_0)
	& \approx &
	- \frac{3}{2} \left( \frac{\nu}{\nu_0} \right)^{p_2 + 1} (p_2 + 2) \int^{\Theta_{\rm bm}}_0 \frac{\Theta^{3}_1 d \Theta_1}{(1 + \Theta^2_1)^{p_2 + 3}} \nonumber \\
	& \approx &
	\frac{3 (p_2 + 2)}{4 (p_2 + 1)} \frac{T_{\rm b}(\nu \Theta^{-2}_{\rm bm})}{T_{\rm b}(\nu_0)},
\end{eqnarray}
where we take $R({\bm e}_z, {\bm \Omega}_1, {\bm u}) \approx 1 + (1 - 2 \Theta^2_1 (1 + \Theta^2_1)^{-1})^2 \approx 1$ because the value varies in the range between $1 \le R({\bm e}_z, {\bm \Omega}_1, {\bm u}) \le 2$ for $0 \le \theta_1 \le \theta_{\rm bm}$.
$I_{\rm Wide}(\nu)$ is order unity at $\nu \sim \Theta^2_{\rm bm} \nu_0$.
Numerical results are shown in Figure \ref{fig:Wide}.
Figure \ref{fig:Wide} shows that $I_{\rm Wide}(\nu)$ is approximated as $- 1$ (order unity) even for $\nu_0 < \nu < \Theta^2_{\rm bm} \nu_0$.
$I_{\rm Wide}(\nu < \nu_0)$ is approximated as $(T(\nu) / T(\nu_0)) S(\nu)$ corresponding to Equation (\ref{eq:NonRelativisticLimit}) with $\theta_{\rm bm} \sim 1$, i.e., almost isotropic.
It is important to note that $I_{\rm Wide} \sim -1$ can be used for applications in Section \ref{sec:Crab} rather than Equation (\ref{eq:WideBeam1}).
Note that $\chi^{}_{\rm Wide}(\nu)$ can also be smaller than $\chi^{}_{\rm NR}(\nu)$ depending on $p_2$ and $\theta_{\rm bm}$ in somewhat complex way because of the frequency shift.

There remains the geometry $\Theta_{\rm bm} > \Theta_{\rm pl} > 1$ where the cone of plasma does not contain ${\bm \Omega}$ but is within the photon beam.
We do not find an analytic form of the integral $I(\nu)$ in this case.
The numerical calculation in Appendix \ref{app:NumericalIntegration} shows that $|I(\nu)|$ takes between $|I_{\rm Inclined}(\nu)|$ and $|I_{\rm Wide}(\nu)|$ for the frequency range $\nu > \nu_0$ in which we are interested in Section \ref{sec:Crab}.
Note that $|I_{\rm Inclined}(\nu)|$ gives the smallest value and $|I_{\rm Wide}(\nu)|$ gives the largest value in any geometries $(\Theta_{\rm bm}, \Theta_{\rm pl})$ for $\nu > \nu_0$.
We give a detailed discussion including this exceptional geometry in Appendix \ref{app:NumericalIntegration}.

\section{APPLICATION TO THE CRAB PULSAR}\label{sec:Crab}

We evaluate the optical depth to induced Compton scattering applying to the Crab pulsar.
We require that the optical depth $|\tau(\nu)|$ is less than unity and then we constrain the Crab pulsar wind properties $\kappa$, $\gamma$, and $\sigma$.

\subsection{Setup}\label{sec:Setup}

We describe assumptions to estimate the normalization $\chi_0$ for the Crab pulsar.
For a pulsar wind, three assumptions are made.
(I) Almost all of the spin-down power $L_{\rm spin}$ goes to the pulsar wind.
(II) The pulsar wind is a cold magnetized $e^{\pm}$ flow whose bulk Lorentz factor is $\gamma$.
(III) The number density of the pulsar wind decreases with $r^{-2}$, and we ignore structures in the pulsar wind, such as the current sheet \cite[e.g.,][]{c90}.
Now, the number density of the pulsar wind in the observer frame is
\begin{eqnarray}\label{eq:Density}
	n_{\rm pl}(r)
	& = &
	\frac{L_{\rm spin}}{4 \pi r^2 c \beta_r \gamma m_{\rm e} c^2 (1 + \sigma)}, \nonumber \\
	& \sim &
	3.2 \times 10^{16} \gamma^{-1} (1 + \sigma)^{-1} \left( \frac{r}{10^8~{\rm cm}} \right)^{-2} \left( \frac{L_{\rm spin}}{10^{38} {\rm erg \cdot s^{-1}}} \right)~{\rm cm}^{-3},
\end{eqnarray}
where we assume the radial velocity $\beta_r \sim 1$.
Note that we obtain Equation (\ref{eq:SpinDownPower}) from Equation (\ref{eq:Density}) by normalizing $4 \pi r^2 c \beta_r n_{\rm pl}(r)$ with $\dot{N}_{\rm GJ}$.
Note also that a product $\gamma (1 + \sigma)$ does not depend on $r$ because we expect no particle production outside the light cylinder $r_{\rm LC}$, i.e., $n_{\rm pl} \propto r^{-2}$.

For radio pulses, uncertainty of the brightness temperature arises from an opening angle of the radio emission $\theta_{\rm bm}$.
Following WR78, we assume that the emission is isotropic at $r = r_{\rm e}$ where $r_{\rm e}$ is an emission region size.
The opening angle $\theta_{\rm bm}(r)$ is written as
\begin{equation}\label{eq:ThetaBeam}
\theta_{\rm bm}(r) \approx \frac{r_{\rm e}}{r} \mbox{ for $r > r_{\rm e}$.}
\end{equation}
We adopt Equation (\ref{eq:ThetaBeam}) for the opening angle of the radio pulse throughout this paper.

The brightness temperature is expressed as \cite[e.g.,][]{lk04}
\begin{eqnarray}\label{eq:BrightnessTemperature}
\frac{k_{\rm B} T_{\rm b}(\nu)}{m_{\rm e} c^2} & = & 1.7 \times 10^{16} \left( \frac{F_{\nu}}{{\rm Jy}} \right) \left( \frac{d}{{\rm kpc}} \right)^2 \left( \frac{\nu}{\rm 100~MHz} \right)^{-2} \left( \frac{r_{\rm e}}{10^7~{\rm cm}} \right)^{-2}, 
\end{eqnarray}
where $F_{\nu}$ and $d$ are a flux density at a frequency $\nu$ and a distance to the object, respectively.
WR78 adopted $r_{\rm e} = 10^7~\rm cm$ which is estimated from the integrated pulse width $W_{50} = 3~\rm msec$ \cite{let95, lk04}.
We study dependence on $r_{\rm e}$ in Section \ref{sec:EmissionRegion}.
In Section \ref{sec:EmissionRegion}, we will take $r_{\rm e} = 10^3$ cm considering the `microbursts' of which individual pulses from the Crab pulsar show nano $-$ microsecond duration structures \cite{he07}.
Note that $r_{\rm e} = 10^3$ cm would also be considered as almost the minimum size of plasma to emit the coherent electromagnetic wave of the frequency $\nu =$ 100 MHz ($c / \nu = 3 \times 10^2$ cm).

\begin{figure}[t]
\centering
\includegraphics[scale=0.7]{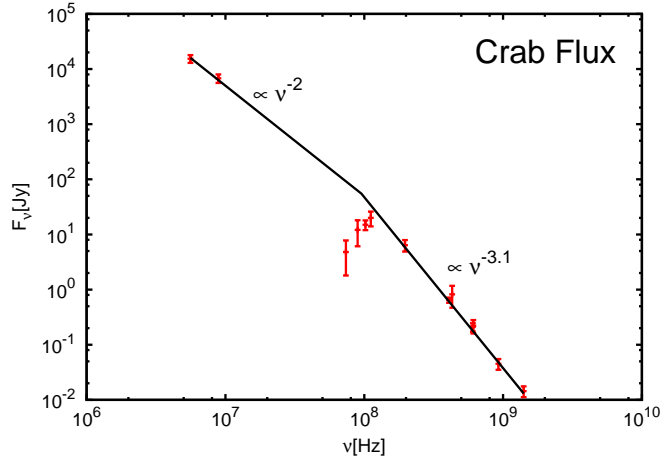}
\caption{
The observed spectrum of the Crab pulsar in radio.
Note that the emission at $\nu < 100~\rm MHz$ is not observed to be pulsed anymore most probably because of the interstellar scattering.
So that apparently rising spectrum around 100 MHz is not real.
Since the high frequency radio flux of the Crab pulsar is $F_{\nu} = 646 (\nu / 400~\rm MHz)^{-3.1}$ mJy for $\nu > 400~\rm MHz$ \cite{let95, met05}, there seems a spectral break around 100 MHz.
The low frequency spectrum extends down to at least 5.6 MHz with a spectral index $\alpha = -2.09$ \cite{ret70, t96}.
Fitted line in this range is $F_{\nu} \sim 50 (\nu / 100~\rm MHz)^{-2}$ Jy for $\nu < 100~\rm MHz$.
Observational data are taken from \cite{ret70, let95, t96}.
}
\label{fig:CrabPulsarFlux}
\end{figure}

Figure \ref{fig:CrabPulsarFlux} shows the radio spectrum of the Crab pulsar.
We assume $F_{\nu} \sim 50~(\nu / 100~{\rm MHz})^{p_2+3}$ Jy for $\nu_0 \le \nu \le 100$ MHz with $\nu_0 = 10$ MHz and $p_2 = -5$.
Adopting $d = 2~\rm kpc$, $L_{\rm spin} = 4.6 \times 10^{38}~\rm erg~s^{-1}$ and the light cylinder radius $r_{\rm LC} = 1.6 \times 10^8~\rm cm$ for the Crab pulsar, we obtain the normalization
\begin{equation}\label{eq:NonRelaCrossSection}
	\chi_{0, \rm Crab}(\nu_0 = 10~{\rm MHz}, r) = 1.3 \times 10^{15} \gamma^{-1} (1 + \sigma)^{-1} \left( \frac{r}{r_{\rm LC}} \right)^{-2} \left( \frac{r_{\rm e}}{10^7~\rm cm} \right)^{-2}.
\end{equation}
Although we used $\nu_0 = 10~\rm MHz$, we require $|\tau(\nu)| < 1$ at $\nu =$ 100 MHz because the Crab pulsar spectrum (Figure \ref{fig:CrabPulsarFlux}) is obviously unaffected by scattering in a range $\nu \ge$ 100 MHz.

On the assumptions made in this section, the scattering coefficient $\chi(\nu, r)$ is considered to be a rapidly decreasing function of $r$.
We introduce the exponents $a$ and $b$ ($(a,~b) > 0$) characterizing the $r$-dependence of the velocity ${\bm u}(r)$ as $\gamma \propto r^a$ and $\theta_{\rm pl} \propto r^{-b}$.
Now, the $r$-dependence of $\chi(\nu, r)$ (Equation (\ref{eq:DefineIntegral})) is expressed as
\begin{eqnarray}\label{eq:RadialDependence}
	\chi_{\rm Narrow}   & \propto & r^{-2} \theta^4_{\rm bm}                                  \propto r^{-6}, \nonumber \\
	\chi_{\rm Inclined} & \propto & r^{-2} \gamma^{-4} \theta^{-4}_{\rm pl} \theta^4_{\rm bm} \propto r^{-6+4(b-a)}, \\
	\chi_{\rm Wide}     & \propto & r^{-2} \gamma^{-4}                                        \propto r^{-2-4a}, \nonumber 
\end{eqnarray}
where $I_{\rm Wide}(\nu) \approx -1$ is used in this section because $\nu_0 \lesssim \nu < \Theta^2_{\rm bm} \nu_0$ ($\nu_0 =$ 10 MHz and $\nu =$ 100 MHz) is mostly attainable for the `Wide' case ($\Theta_{\rm bm} > 1$).
In Equation (\ref{eq:RadialDependence}), $b-a < 1.25$ is sufficient for $\chi(\nu, r)$ to be considered as a rapidly decreasing function of $r$.
Otherwise we consider moderate values of $a$ and $b$, say, $0 < (a,~b) \lesssim 1.25$ below.
Therefore, the choice of the innermost scattering radius is important to evaluate the optical depth.

Here, we consider scattering beyond the light cylinder $r \ge r_{\rm LC}$, because we do not know where the electron-positron plasma and the radio emission are produced inside the magnetosphere and because we do not take into account magnetic field effects which may be important close to the pulsar.
We evaluate the optical depth as
\begin{eqnarray}\label{eq:OpticalDepthIntegral}
\tau(\nu) & =    & \int^{d}_{r_{\rm in}} \chi(\nu, r) d r \nonumber \\
	  & \sim & \chi(\nu, r = r_{\rm in}) \Delta r,
\end{eqnarray}
where $r_{\rm in}$ and $\Delta r$ are the innermost scattering radius and the path length, respectively.
In Equation (\ref{eq:OpticalDepthIntegral}), we should not simply put $r_{\rm in} = \Delta r = r_{\rm LC}$ because the path length $\Delta r$ has a lower limit originating from the `lack of time' effect which we will discuss in the next subsection.

\subsection{Characteristic Scattering Length}\label{sec:CharacteristicScatteringLength}

The `lack of time' effect introduced by WR78 should be taken into account for the evaluation of $r_{\rm in}$ and $\Delta r$ in Equation (\ref{eq:OpticalDepthIntegral}).
This is similar to the concept of the `coherence radiation length' \cite[e.g.,][]{gg64, as87}.
The normal treatment of scattering breaks down when an electron does not see one cycle of the electric field oscillation of radio waves.
We determine this characteristic length $l_{\rm c}$ as follows.
A cycle of the incident and scattered photons in the plasma rest frame is described as $\Delta t' = \delta_{\rm D} / \nu$ where $\delta_{\rm D} = (\gamma D)^{-1}$ or $(\gamma D_1)^{-1}$ is the Doppler factor.
The characteristic length $l_{\rm c}$ is the speed of light multiplied by the time interval $\Delta t = \gamma \Delta t'$ in the observer frame.
Using $D^{-1} \approx 2 \gamma^2 / (1 + \Theta^2_{\rm pl})$ and $D^{-1}_1 \approx 2 \gamma^2 / (1 + \Psi^2_1)$ ($\Psi^2_1 \equiv \gamma^2 \psi^2_1$), we obtain
\begin{eqnarray}\label{eq:CharacteristicRadius}
	l_{\rm c}(\nu, {\bm u}, {\bm \Omega}, {\bm \Omega}_1) 
	& = & 
	\frac{c}{\nu} \max(D^{-1}, D^{-1}_1) \nonumber \\
	& \approx &
	2 \gamma^2 \frac{c}{\nu} \times \left\{
	\begin{array}{ll}
		\max(1                   , 1                   ) & \mbox{ for `Narrow',} \\
		\max(\Theta^{-2}_{\rm pl}, \Theta^{-2}_{\rm pl}) & \mbox{ for `Inclined',} \\
		\max(1                   , (1 + \Psi^2_1)^{-1} ) & \mbox{ for `Wide',} \\
	\end{array} 
	\right.
\nonumber  \\
	& = &
	6 \times 10^2~{\rm cm}
	\left( \frac{\nu}{100~{\rm MHz}} \right)^{-1} \times \left\{
	\begin{array}{ll}
	\gamma^2             & \mbox{ for `Narrow' and `Wide',} \\
	\theta^{-2}_{\rm pl} & \mbox{ for `Inclined'.} \\
	\end{array} 
	\right.
\end{eqnarray}
$l_{\rm c}$ is considered as a function of only $r$ through $\gamma(r)$ or $\theta_{\rm pl}(r)$ for the given frequency $\nu = 100~{\rm MHz}$.
On the other hand, for the geometry $\Theta_{\rm bm} > \Theta_{\rm pl} > 1$, we obtain
\begin{eqnarray}\label{eq:CharacteristicRadius2}
	\max(D^{-1}, D^{-1}_1)
	& \approx &
	\left\{
	\begin{array}{ll}
		D^{-1}   & \mbox{ for $\Theta^2_{\rm pl} \le \Psi^2_1$,} \\
		D^{-1}_1 (> D^{-1}) & \mbox{ for $\Theta^2_{\rm pl} >   \Psi^2_1$,} \\
	\end{array} 
	\right. \nonumber \\
	& \ge &
	D^{-1}
	\approx
	2 \theta^{-2}_{\rm pl}.
\end{eqnarray}
%
We find $l_{\rm c}$ for this case is equal to or larger than that for the `Inclined' case.
Because $l_{\rm c}$ depends on ${\bm \Omega}_1$, we cannot separate integrals over ${\bm \Omega}_1$ and $r$ in Equations (\ref{eq:DefineIntegral}) and (\ref{eq:OpticalDepthIntegral}).
In this subsection, we limit the discussion about the `Narrow', `Inclined' and `Wide' cases.

\begin{figure}[t]
\centering
\includegraphics[scale=0.4]{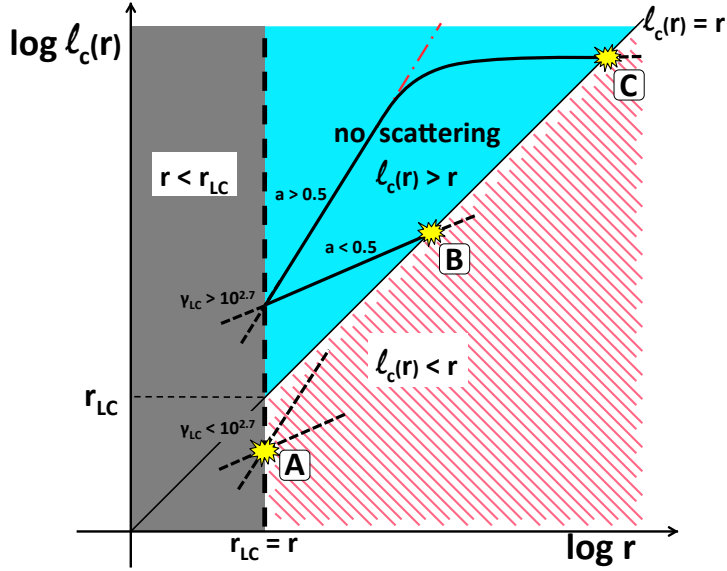}
\caption{
	The $l_{\rm c} - r$ diagram for the `Narrow' and `Wide' cases (Equation (\ref{eq:RadialDepndenceOfCharacteristicRadius})).
	The region $r < r_{\rm LC}$ (grey in color) is not considered in this paper.
	When $l_{\rm c}(r) > r$ (light blue in color), scattering does not occur because of the `lack of time' effect, while scattering should be considered in the region $l_{\rm c}(r) < r$ (pink in color).
	Three cases for $r_{\rm in}$ (points `A', `B' and `C') are possible by different behaviors of $l_{\rm c}(r)$, i.e., $\gamma_{\rm LC}$ and the exponent $a$ (see also Equation (\ref{eq:RadialDepndenceOfCharacteristicRadius})).
	$r_{\rm in}$ becomes point `A' when $\gamma_{\rm LC} < 10^{2.7}$.
	For $\gamma_{\rm LC} > 10^{2.7}$, $r_{\rm in}$ is point `B' when $a < 0.5$.
	While no $r_{\rm in}$ exists for $a > 0.5$, because $\gamma$, i.e., $l_{\rm c}(r)$, cannot be infinitely large, there must be point `C' where $l_{\rm c}(r) = r$ is satisfied.
}
\label{fig:lc_r}
\end{figure}

Now, we describe how we determine $r_{\rm in}$ and $\Delta r$ taking into account the $r$-dependence of $l_{\rm c}(r)$.
Although we describe only for the `Narrow' and `Wide' cases ($l_{\rm c} \propto \gamma^2$), the same discussion is applicable to the `Inclined' case ($l_{\rm c} \propto \theta^{-2}_{\rm pl}$) by replacing $\gamma$ with $\theta^{-1}_{\rm pl}$.
We set $\gamma(r) = \gamma_{\rm LC} (r/r_{\rm LC})^a$ where $\gamma_{\rm LC}$ is the Lorentz factor at $r_{\rm LC}$.
Substituting it into Equation (\ref{eq:CharacteristicRadius}), we obtain
\begin{equation}\label{eq:RadialDepndenceOfCharacteristicRadius}
	l_{\rm c}(r) = 6 \times 10^2~\gamma^2_{\rm LC} \left( \frac{r}{r_{\rm LC}} \right)^{2a}~{\rm cm}.
\end{equation}
In Figure \ref{fig:lc_r}, we show the $l_{\rm c}(r) - r$ diagram.
We do not consider the region $r < r_{\rm LC}$.
The region $r > r_{\rm LC}$ is divided into two regions by the line $l_{\rm c}(r) = r$ which corresponds to $\gamma_{\rm LC} \approx 10^{2.7}$ and $a = 0.5$.
Scattering off the radio pulse should be considered when $l_{\rm c}(r) < r$ so that three different choices of $r_{\rm in}$ are possible for different values of $\gamma_{\rm LC}$ and the exponent $a$, corresponding to points `A', `B' and `C' in Figure \ref{fig:lc_r}.
Point `A' corresponds to $\gamma_{\rm LC} < 10^{2.7}$ with any values of the exponent $a$. 
Since $l_{\rm c}(r_{\rm LC}) < r_{\rm LC}$ in this case, we take $r_{\rm in} = \Delta r = r_{\rm LC}$.
Point `B' corresponds to $\gamma_{\rm LC} > 10^{2.7}$ with $a < 0.5$.
The radio pulse is not scattered at $r_{\rm LC}$ but beyond $r_{\rm LC}$.
Here, we introduce the characteristic scattering radius $r_{\rm c}$ which satisfies $r_{\rm c} = l_{\rm c}(r_{\rm c}) > r_{\rm LC}$ so that we take $r_{\rm in} = \Delta r = r_{\rm c} = (10^{2.8} \gamma^{2}_{\rm LC} r^{-2a}_{\rm LC})^{1/(1-2a)}$ cm.
For $\gamma_{\rm LC} > 10^{2.7}$ with $a \ge 0.5$, we obtain $l_{\rm c}(r) > r$ everywhere beyond $r_{\rm LC}$, i.e., the electron never sees one cycle of radio waves (dot-dashed line: red in color).
However, $\gamma(r)$ cannot be infinitely large so that there should exist the radius satisfying $r_{\rm in} = l_{\rm c}(r_{\rm in}) > r_{\rm LC}$ corresponding to point `C'.
In this case, we also take $r_{\rm in} = \Delta r = r_{\rm c}$ whose expression is different from that for $a < 0.5$.
Therefore, $\gamma_{\rm LC} = 10^{2.7}$ or $\theta_{\rm pl, LC} = 10^{-2.7}$ is a critical value in determining which to adopt as $r_{\rm in}$.

We consider whether the radio pulse can escape from scattering at the two radii $r_{\rm LC}$ and $r_{\rm c}$.
Rather than using the exponents $a$ and/or $b$, it is convenient to introduce $\gamma_{\rm c} \equiv \gamma(r_{\rm c})$ and $\theta_{\rm pl, c} \equiv \theta_{\rm pl}(r_{\rm c})$.
We evaluate the optical depth by treating the velocities ${\bm u}_{\rm LC}$ and ${\bm u}_{\rm c}$, i.e., ($\gamma_{\rm LC}$, $\theta_{\rm pl, LC}$) and ($\gamma_{\rm c}$, $\theta_{\rm pl, c}$), as free parameters.
Relation between the exponent $a$ ($b$) and $\gamma_{\rm c}$ ($\theta_{\rm pl, c}$) will be discussed shortly in Section \ref{sec:escapesummary}.
Note that we indirectly obtain the characteristic scattering radius $r_{\rm c}$ from Equation (\ref{eq:CharacteristicRadius}) once $\gamma_{\rm c}$ or $\theta_{\rm pl, c}$ is obtained.

\subsection{Constrains on Lorentz Factor}\label{sec:LorentzFactor}
%
Lower limits of $\gamma$ are obtained from the condition $|\tau(\nu)| < 1$ for a given $\theta_{\rm pl}$.
We evaluate the optical depth,
\begin{eqnarray}\label{eq:CrabOpticalDepth}
\tau(\nu) \sim \Delta r \chi_{\rm 0, Crab}(r_{\rm in}) \gamma^{-3} I(\nu, \theta_{\rm bm}, \theta_{\rm pl}, \gamma),
\end{eqnarray}
at $\nu = 100$ MHz.
$\tau(100~{\rm MHz})$ strongly depends on ${\bm u}_{\rm LC}$ or ${\bm u}_{\rm c}$ (Tables \ref{tbl:1e7rlc} and \ref{tbl:1e7rc}).
Below, we search allowable region on $\gamma - \theta_{\rm pl}$ planes for $r_{\rm in} = r_{\rm LC}$ (Figure \ref{fig:ConstraintAtRlc1e7}) and for $r_{\rm in} = r_{\rm c}$ (Figure \ref{fig:ConstraintAtRc1e7}), respectively.
The results will be combined in Section \ref{sec:escapesummary}.

For a given $r_{\rm e}$, i.e., $\theta_{\rm bm}(r_{\rm in})$ (Equation (\ref{eq:ThetaBeam})), scattering geometry is classified into four cases on the $\gamma - \theta_{\rm pl}$ plane corresponding to the `Narrow' ($1 > \Theta^2_{\rm bm} + \Theta^2_{\rm pl}$), `Inclined' ($\Theta^2_{\rm pl} > \Theta^2_{\rm bm} + 1$) and `Wide' ($\Theta_{\rm bm} > 1 > \Theta_{\rm pl}$) cases, and the geometry satisfying $\Theta_{\rm bm} > \Theta_{\rm pl} > 1$.
The first three geometries are studied in section \ref{sec:AnalyticEstimates} and the expressions of $\tau(100~{\rm MHz})$ for them are obtained in Tables \ref{tbl:1e7rlc} and \ref{tbl:1e7rc}.
For $\Theta_{\rm bm} > \Theta_{\rm pl} > 1$, $\tau(100~{\rm MHz})$ is not expressed by Equation (\ref{eq:CrabOpticalDepth}) because $l_{\rm c}$ depends on ${\bm \Omega}_1$ as already discussed in Equation (\ref{eq:CharacteristicRadius2}).
Here, we infer the optical depth for $\Theta_{\rm bm} > \Theta_{\rm pl} > 1$ from the resuls of other three cases.
Thus, the $|\tau(100~{\rm MHz})| = 1$ lines at the $\Theta_{\rm bm} > \Theta_{\rm pl} > 1$ area in Figures \ref{fig:ConstraintAtRlc1e7} and \ref{fig:ConstraintAtRc1e7} (thick dashed lines) are not calculated but inferred ones.

We adopt $r_{\rm e} = 10^7 r_{\rm e, 7}$ cm to evaluate $\theta_{\rm bm}(r_{\rm in})$ and will study when $r_{\rm e} = 10^3$ cm in Section \ref{sec:EmissionRegion} ($r_{\rm e}$-dependence is already included explicitly in Tables \ref{tbl:1e7rlc} and \ref{tbl:1e7rc}).
We consider customarily used values of $\sigma$ ($\sigma_{\rm LC}$ and $\sigma_{\rm c}$) in a range of $1 < 1 + \sigma \lesssim 10^4$.
We take $\nu_0 =$ 10 MHz, $\nu =$ 100 MHz and $p_2 = -5$, i.e., $T_{\rm b}(\nu) / T_{\rm b}(\nu_0) \sim 10^{-4}$ in the integrals $I_{\rm Narrow}(\nu)$ and $I_{\rm Inclined}(\nu)$
While $I_{\rm Wide}(\nu) \sim - 1$ is used as the same reason discussed in Equation (\ref{eq:RadialDependence}).
Again, only the pulsar wind velocities ${\bm u}_{\rm LC}$ and ${\bm u}_{\rm c}$ are remaining parameters, i.e., we take ($\gamma_{\rm LC}$, $\theta_{\rm pl, LC}$) and ($\gamma_{\rm c}$, $\theta_{\rm pl, c}$) as the free parameters.

\subsubsection{Escape from scattering at the light cylinder}\label{sec:escape1}
\begin{figure}[t]
\centering
\includegraphics[scale=0.4]{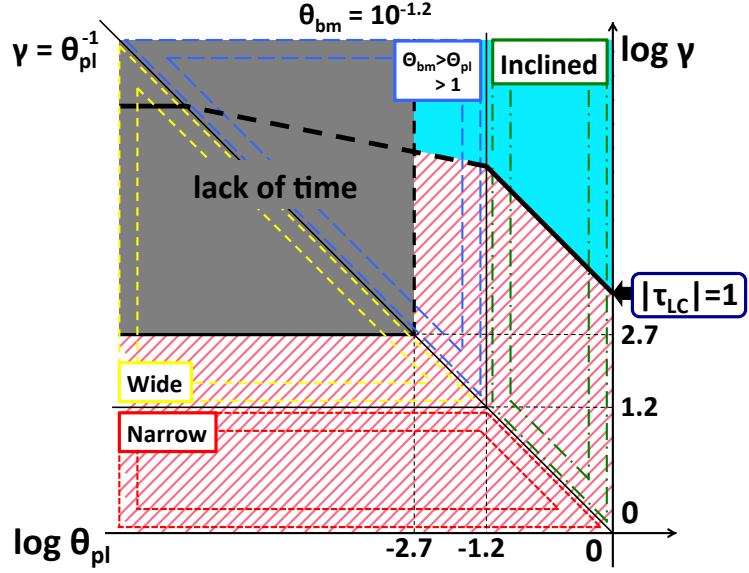}
\caption{
	The $\gamma - \theta_{\rm pl}$ diagram at $r_{\rm LC}$ when $r_{\rm e} = 10^7$ cm ($\theta_{\rm bm}(r_{\rm LC}) \approx 10^{-1.2}$).
	Choosing one point on the diagram specifies the pulsar wind velocity ${\bm u}_{\rm LC}$.
	Four areas divided by three lines $\gamma = \theta^{-1}_{\rm pl}$, $\gamma = 10^{1.2}$ and $\theta_{\rm pl} = 10^{-1.2}$ correspond to different scattering geometries, the `Narrow' (lowermost area: red in color), `Inclined' (rightmost area: green in color) and `Wide' (left triangle area: yellow in color) cases and the geometry $\Theta_{\rm bm} > \Theta_{\rm pl} > 1$ (upper triangle area: blue in color).
	The region above the $|\tau_{\rm LC}| = 1$ line (light blue in color) corresponds to $|\tau_{\rm LC}| < 1$, i.e., where the pulsar wind does not scatter the radio pulses at and beyond $r_{\rm LC}$.
	The upper left corner which satisfies $\gamma > 10^{2.7}$ and $\theta_{\rm pl} < 10^{-2.7}$ (gray in color) corresponds to $l_{\rm c}(r_{\rm LC}) > r_{\rm LC}$, i.e., the radio pulses are not scattered at $r_{\rm LC}$ because of the `lack of time' effect and we also require $|\tau_{\rm c}| < 1$ in Figure \ref{fig:ConstraintAtRc1e7}.
	The $|\tau_{\rm LC}| = 1$ lines (thick lines) at the `Inclined' and `Wide' areas are determined by $\gamma_{\rm LC} = 10^{3.7} \theta^{-1}_{\rm pl, LC} (1 + \sigma_{\rm LC})^{-1/4}$
	and $\gamma_{\rm LC} = 10^{5.8} (1 + \sigma_{\rm LC})^{-1/4}$, respectively and depend on $\sigma_{\rm LC}$ (see also Table \ref{tbl:1e7rlc}).
	We adopt $1 < 1 + \sigma_{\rm LC} \ll 10^4$ in the diagram, for example, y-intercept of the $|\tau_{\rm LC}| = 1$ line in the `Inclined' area is $\gamma \sim 10^{3.2}$ for $1 + \sigma_{\rm LC} \sim 10^2$.
	Note that the line in the geometry $\Theta_{\rm bm} > \Theta_{\rm pl} > 1$ is drawn in a dashed line because it is an interpolated ones (see text).
	On the other hand, the shaded region (pink in color) is the forbidden region.
}
\label{fig:ConstraintAtRlc1e7}
\end{figure}

%
Here, we are interested in whether the radio pulse can escape from scattering at $r_{\rm LC}$.
Figure \ref{fig:ConstraintAtRlc1e7} shows the resultant $\gamma - \theta_{\rm pl}$ diagram which tells us whether the radio pulses can escape from scattering or not at a given point on the diagram, i.e., a given velocity ${\bm u}_{\rm LC}$ of the pulsar wind (see also Table \ref{tbl:1e7rlc}).
Since we obtain $\theta_{\rm bm}(r_{\rm LC}) \approx 10^{-1.2}$ from Equation (\ref{eq:ThetaBeam}), the scattering geometries are divided by the lines $\gamma = 10^{1.2}$ ($\Theta_{\rm bm, LC} = 1$), $\theta_{\rm pl} = 10^{-1.2}$ ($\Theta_{\rm pl} = \Theta_{\rm bm, LC}$) and $\gamma = \theta^{-1}_{\rm pl}$ ($\Theta_{\rm pl} = 1$).
Areas above the thick lines $|\tau_{\rm LC}| = 1$ correspond to the pulsar wind structures which allow the radio pulses to escape, where $\tau_{\rm LC}$ is the optical depth for $r_{\rm in} = r_{\rm LC}$. 
At the upper left corner on the diagram, the region satisfies $l_{\rm c}(r_{\rm LC}) > r_{\rm LC}$ and the radio pulses also escape from scattering at $r_{\rm LC}$ due to the `lack of time' effect.
The lines $|\tau_{\rm LC}| = 1$ and $l_{\rm c}(r_{\rm LC}) = r_{\rm LC}$ are different for different scattering geometries as described below and summarized in Table \ref{tbl:1e7rlc}.

\begin{table}[!t]
\begin{minipage}{1.0\hsize}
\caption{
	The optical depth $|\tau({\rm 100~MHz})|$ at $r_{\rm LC}$ ($\gamma_{\rm LC} < 10^{2.7}$ or $\theta_{\rm pl, LC} > 10^{-2.7}$).
	Scattering geometries are classified by ${\bm u}_{\rm LC}$, i.e., $\gamma_{\rm LC}$ and $\theta_{\rm pl, LC}$.
	We take $r_{\rm e} = 10^7 r_{\rm e, 7}$ cm.
}
\label{tbl:1e7rlc}
\begin{center}
\begin{tabular}{ccc}
\hline
Geometry              &
$\tau({\rm 100~MHz})$ &
${\bm u}(r_{\rm LC})$ \\
\hline
`Narrow'                                 &
$10^{14.9} r^2_{\rm e, 7} (1 + \sigma_{\rm LC})^{-1}$ 
&
$\gamma_{\rm LC} \lesssim 10^{1.2} r^{-1}_{\rm e, 7}$ \& $\gamma_{\rm LC} < \theta^{-1}_{\rm pl, LC}$ \\
`Inclined'                                               &
$10^{14.9} r^2_{\rm e, 7} \gamma^{-4}_{\rm LC} \theta^{-4}_{\rm pl, LC} (1 + \sigma_{\rm LC})^{-1}$  
& 
$\theta_{\rm pl, LC} \gtrsim 10^{-1.2} r_{\rm e, 7}$ \& $\gamma_{\rm LC} > \theta^{-1}_{\rm pl, LC}$ \\
`Wide'                                               &
$-10^{23.3} r^{-2}_{\rm e, 7} \gamma^{-4}_{\rm LC} (1 + \sigma_{\rm LC})^{-1}$  & 
$\gamma_{\rm LC} \gtrsim 10^{1.2} r^{-1}_{\rm e, 7}$ \& $\gamma_{\rm LC} < \theta^{-1}_{\rm pl, LC}$ \\
\hline
\end{tabular}
\end{center}
\end{minipage}
\end{table}

First, we consider the `Narrow' case ($1 > \Theta^2_{\rm bm} + \Theta^2_{\rm pl}$) corresponding to the lowermost area on the diagram.
The optical depth of $\tau_{\rm LC} \sim 10^{14.9} (1 + \sigma_{\rm LC})^{-1}$ obtained from Equations (\ref{eq:NarrowBeam}), (\ref{eq:NonRelaCrossSection}) and (\ref{eq:CrabOpticalDepth}) is independent of both $\gamma_{\rm LC}$ and $\theta_{\rm pl, LC}$.
Therefore, a region $|\tau_{\rm LC}| < 1$ does not appear for $1 + \sigma_{\rm LC} \lesssim 10^4$ and then we conclude that this case is not realized for the Crab pulsar.

Next, we consider the `Inclined' case ($\Theta^2_{\rm pl} > \Theta^2_{\rm bm} + 1$) corresponding to the rightmost area on the diagram.
In this case, the optical depth is expressed as $\tau_{\rm LC} \sim 10^{14.9} \gamma^{-4}_{\rm LC} \theta^{-4}_{\rm pl, LC} (1 + \sigma_{\rm LC})^{-1}$. 
The condition for $|\tau_{\rm LC}| < 1$ is equivalent to $\gamma_{\rm LC} \gtrsim 10^{3.7} \theta^{-1}_{\rm pl, LC} (1 + \sigma_{\rm LC})^{-1/4}$ with $\theta_{\rm pl, LC} \gtrsim 10^{-1.2}$ where the painted area above $|\tau_{\rm LC}| = 1$ line in the `Inclined' area on the diagram.
We find that the radio pulses can escape for reasonable parameters when the pulsar wind has a significant non-radial motion.
For example, the pulsar wind of $\gamma_{\rm LC} > 10^{2.7}$ with $\theta_{\rm pl, LC} \sim 1$ and $1 + \sigma_{\rm LC} \approx 10^4$ can escape from scattering at $r_{\rm LC}$.

The `Wide' case ($\Theta_{\rm bm} > 1 > \Theta_{\rm pl}$) corresponds to the left triangle area on the diagram.
For $|\tau_{\rm LC}| \sim 10^{23.3} \gamma^{-4}_{\rm LC} (1 + \sigma_{\rm LC})^{-1}$ to be less than unity, we require $\gamma_{\rm LC} > 10^{5.8} (1 + \sigma_{\rm LC})^{-1/4}$ where the $|\tau_{\rm LC}| = 1$ line in the `Wide' area on the diagram.
However, because the line is already above $\gamma_{\rm LC} > 10^{2.7}$ for $1 < 1 + \sigma_{\rm LC} \lesssim 10^4$, therefore, $\gamma_{\rm LC} > 10^{2.7}$ (the `lack of time' effect) is the condition for the radio pulses to escaping from scattering at $r_{\rm LC}$ in this case.

Lastly, we mention the geometry of $\Theta_{\rm bm} > \Theta_{\rm pl} > 1$ which appears in the upper triangle area on the diagram.
The $l_{\rm c}(r_{\rm LC}) = r_{\rm LC}$ and $|\tau_{\rm LC}| = 1$ lines (dashed lines) are not calculated but interpolated ones.
For escaping by the `lack of time' effect ($l_{\rm c}(r_{\rm LC}) > r_{\rm LC}$), we obtain at least $\theta_{\rm pl, LC} < 10^{-2.7}$ from Equation (\ref{eq:CharacteristicRadius2}).
The $|\tau_{\rm LC}| = 1$ line is expected to be continuous at the boundaries on the $\gamma_{\rm LC} = \theta^{-1}_{\rm pl, LC}$ and $\theta_{\rm pl, LC} = 10^{-1.2}$ lines because these boundaries just divide the approximated forms of Equation (\ref{eq:DefineIntegral}).
On the other hand, the $|\tau_{\rm LC}| = 1$ line would have at least one singular point because $\tau_{\rm LC}$ changes the sign at the left and right boundaries and a singular line (or curve) which satisfies $\tau_{\rm LC} = 0$ would be drawn on the diagram.
Although a significantly small value of $\gamma_{\rm LC}$ might	be allowed on the sides of the singular line, such a region on the $\gamma - \theta_{\rm pl}$ diagram would be as small as the dip around the discontinuity of $I(\nu)$ in Figures \ref{fig:Narrow} $-$ \ref{fig:Wide} because $S(\nu_1)$ which appears in Equation (\ref{eq:DefineIntegral}) controls the singularity $\tau_{\rm LC} = 0$.
When we neglect such a singular region, the allowed region would be above the thick dashed line and the lower limit of $\gamma_{\rm LC}$ is clearly larger than the `Inclined' case.

\subsubsection{Escape from scattering beyond the light cylinder}\label{sec:escape2}
\begin{figure}[t]
\centering
\includegraphics[scale=0.4]{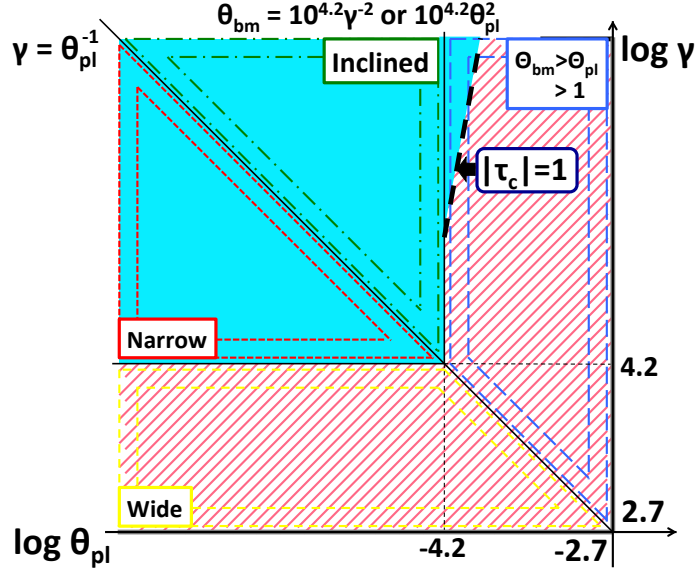}
\caption{
	The $\gamma - \theta_{\rm pl}$ diagram at $r_{\rm c}$ ($> r_{\rm LC}$) when $r_{\rm e} = 10^7$ cm ($\theta_{\rm bm}(r_{\rm c}) \approx 10^{4.2} \gamma^{-2}_{\rm c}$ or $10^{4.2} \theta^2_{\rm pl, c}$).
	We show only the region which satisfies both $\gamma > 10^{2.7}$ and $\theta_{\rm pl} < 10^{-2.7}$ because we consider the case $r_{\rm c} > r_{\rm LC}$.
	The pulsar wind velocity ${\bm u}_{\rm c}$ is specified by choosing one point on the diagram.
	Four areas divided by three lines $\gamma = \theta^{-1}_{\rm pl}$, $\gamma = 10^{4.2}$ and $\theta_{\rm pl} = 10^{-4.2}$ correspond to different scattering geometries, the `Narrow' (left triangle area: red in color), `Inclined' (upper triangle area: green in color) and `Wide' (lowermost area: yellow in color) cases and the geometry $\Theta_{\rm bm} > \Theta_{\rm pl} > 1$ (rightmost area: blue in color).
	Note that each scattering geometry appears in a different layout compared with Figure \ref{fig:ConstraintAtRlc1e7}.
	The painted region (light blue in color) satisfies $|\tau_{\rm c}| < 1$, i.e., the radio pulses are not scattered at $r_{\rm c} \gtrsim 10^{11.2}~{\rm cm} \approx 10^3 r_{\rm LC}$.
	The $|\tau_{\rm c}| = 1$ line (dashed thick line) appears only in the geometry $\Theta_{\rm bm} > \Theta_{\rm pl} > 1$ and is an extrapolated one (see text).
	On the other hand, the shaded region (pink in color) is forbidden region because $|\tau_{\rm c}| > 1$ or, in other words, $r_{\rm c} < 10^{11.2}$ cm.
}
\label{fig:ConstraintAtRc1e7}
\end{figure}

%
We investigate whether the radio pulse can escape from scattering at $r_{\rm c}$ further than $r_{\rm LC}$.
Because $r_{\rm c} > r_{\rm LC}$, we have only to consider a region of $\gamma > 10^{2.7}$ and $\theta_{\rm pl} < 10^{-2.7}$.
The behaviors of $\gamma(r)$ and $\theta_{\rm pl}(r)$ at $r_{\rm LC} < r < r_{\rm c}$ will be discussed in Section \ref{sec:escapesummary}.
Figure \ref{fig:ConstraintAtRc1e7} shows the resultant $\gamma - \theta_{\rm pl}$ diagram at $r_{\rm c}$.
We set $\theta_{\rm bm}(r_{\rm c}) \approx 10^{4.2} \gamma^{-2}_{\rm c}$ for the `Narrow' and `Wide' cases or $\theta_{\rm bm}(r_{\rm c}) \approx 10^{4.2} \theta^2_{\rm pl, c}$ for the `Inclined' case from Equations (\ref{eq:ThetaBeam}) and (\ref{eq:CharacteristicRadius}).
The scattering geometries are divided by the lines $\gamma = 10^{4.2}$ ($\Theta_{\rm bm, c} = 1$), $\theta_{\rm pl} = 10^{-4.2}$ ($\Theta_{\rm pl} = \Theta_{\rm bm, c}$) and $\gamma = \theta^{-1}_{\rm pl}$ ($\Theta_{\rm pl} = 1$) (see Table \ref{tbl:1e7rc}).
It should be noted that each scattering geometry appears in a different layout on the $\gamma - \theta_{\rm pl}$ diagram compared with Figure \ref{fig:ConstraintAtRlc1e7} because $\theta_{\rm bm}(r_{\rm c})$ depends on $\gamma_{\rm c}$ or $\theta_{\rm pl, c}$.
The pulsar wind velocity ${\bm u}_{\rm c}$ which allows the radio pulses to escape corresponds to the area satisfying $\gamma_{\rm c} \ge 10^{4.2}$ and $\theta_{\rm pl, c} \le 10^{-4.2}$ corresponding to the `Narrow' or `Inclined' cases.
Except for the extrapolated line in the geometry $\Theta_{\rm bm} > \Theta_{\rm pl} > 1$ (thick dashed line), the $|\tau_{\rm c}| = 1$ line is not drawn on the diagram as described below, where $\tau_{\rm c}$ is the optical depth for $r_{\rm in} = r_{\rm c}$.

\begin{table}[!t]
\begin{minipage}{1.0\hsize}
\caption{
	The optical depth $|\tau({\rm 100~MHz})|$ at $r_{\rm c}$ ($\gamma_{\rm LC} > 10^{2.7}$ and $\theta_{\rm pl, LC} < 10^{-2.7}$).
	Scattering geometries are classified by ${\bm u}_{\rm c}$.
	We take $r_{\rm e} = 10^7 r_{\rm e, 7}$ cm.
}
\label{tbl:1e7rc}
\begin{center}
\begin{tabular}{cccc}
\hline
Geometry &
$\tau({\rm 100~MHz})$    &
${\bm u}(r_{\rm c})$  \\        
\hline
`Narrow'                                 &
$10^{42.0} r^2_{\rm e, 7} \gamma^{-10}_{\rm c} (1 + \sigma_{\rm c})^{-1}$
& 
$\gamma_{\rm c} \gtrsim 10^{4.2} r_{\rm e, 7}$ \& $\gamma_{\rm c} < \theta^{-1}_{\rm pl, c}$ \\
`Inclined'                                               &
$10^{42.0} r^2_{\rm e, 7} \gamma^{-4}_{\rm c} \theta^6_{\rm pl, c} (1 + \sigma_{\rm c})^{-1}$  
& 
$\theta_{\rm pl, c} \lesssim 10^{-4.2} r^{-1}_{\rm e, 7}$ \& $\gamma_{\rm c} > \theta^{-1}_{\rm pl, c}$ \\
`Wide'                                               &
$-10^{28.7} r^{-2}_{\rm e, 7} \gamma^{-6}_{\rm c} (1 + \sigma_{\rm c})^{-1}$    & 
$\gamma_{\rm c} \lesssim 10^{4.2} r_{\rm e, 7}$ \& $\gamma_{\rm c} < \theta^{-1}_{\rm pl, c}$ \\
\hline
\end{tabular}
\end{center}
\end{minipage}
\end{table}

The `Narrow' case ($1 > \Theta^2_{\rm bm} + \Theta^2_{\rm pl}$) corresponds to the left triangle area on the diagram.
In this case, the optical depth is written as $\tau_{\rm c} \approx 10^{42.0} \gamma^{-10}_{\rm c} (1 + \sigma_{\rm c})^{-1}$, i.e., we require $\gamma_{\rm c} \gtrsim 10^{4.2} (1 + \sigma_{\rm c})^{-1/10}$ to be $|\tau_{\rm c}| < 1$.
The $|\tau_{\rm c}| = 1$ line is degenerate to or a bit lower than the $\gamma_{\rm c} = 10^{4.2}$ line for $1 + \sigma_{\rm c} > 1$.
Therefore, whole of the `Narrow' geometry area $\gamma_{\rm c} \ge 10^{4.2}$ is allowed for radio pulses to escape.
The corresponding characteristic scattering radius is $r_{\rm c} \gtrsim 10^{11.2}~{\rm cm} \sim 10^3 r_{\rm LC}$.

Next, we consider the `Inclined' case ($\Theta^2_{\rm pl} > \Theta^2_{\rm bm} + 1$) corresponding to the right triangle area on the diagram.
For the optical depth, we require $|\tau_{\rm c}| \sim 10^{42.0} \gamma^{-4}_{\rm c} \theta^6_{\rm pl, c} (1 + \sigma_{\rm c})^{-1} = 10^{42.0} \gamma^{-10}_{\rm c} \Theta^6_{\rm pl, c} (1 + \sigma_{\rm c})^{-1} < 1$ at $r_{\rm c}$.
The $|\tau_{\rm c}| = 1$ line satisfies $\gamma_{\rm c} = 10^{4.2} \Theta^{3/5}_{\rm pl, c} (1 + \sigma_{\rm c})^{-1/10}$ which has slope $\gamma \propto \theta^{3/2}_{\rm pl}$ and is continuous with the $|\tau_{\rm c}| = 1$ line for the `Narrow' case on the boundary line $\gamma = \theta^{-1}_{\rm pl}$.
Note that large $\theta_{\rm pl, c}$ does not reduce $|\tau_{\rm c}|$ as $|\tau_{\rm LC}|$ is reduced by large $\theta_{\rm pl, LC}$ (see the `Inclined' area in Figure \ref{fig:ConstraintAtRlc1e7}) because $r_{\rm c}$ is a rapidly decreasing function of $\theta_{\rm pl, c}$.
Therefore, whole of the `Inclined' geometry area $\theta_{\rm pl, c} \le 10^{-4.2}$ is allowed for radio pulses to escape and we obtain $r_{\rm c} \gtrsim 10^{11.2}$ cm again.

The `Wide' case ($\Theta_{\rm bm} > 1 > \Theta_{\rm pl}$) corresponding to the lowermost area on the diagram. 
The condition to be $|\tau_{\rm c}| < 1$ is $\gamma_{\rm c} \gtrsim 10^{4.8} (1 + \sigma_{\rm c})^{-1/6}$.
In this case, a region $|\tau_{\rm c}| < 1$ does not appear in the `Wide' area for $1 + \sigma_{\rm c} < 10^{4}$ and then we conclude that this case is not realized for the Crab pulsar.

For the geometry of $\Theta_{\rm bm} > \Theta_{\rm pl} > 1$ corresponding to the rightmost area on the diagram, we do not draw the $|\tau_{\rm c}| = 1$ line in the same manner as Figure \ref{fig:ConstraintAtRlc1e7} because no $|\tau_{\rm c}| = 1$ line appears in Figure \ref{fig:ConstraintAtRc1e7} for other geometries.
One possibility is that the $|\tau_{\rm c}| = 1$ line emerges from the boundary $\theta_{\rm pl} = 10^{-4.2}$, such as the thick dashed line on the diagram.
As implied from the $|\tau_{\rm c}| = 1$ line for the `Inclined' case, the line has slope $\gamma \propto \theta^{q}_{\rm pl}$ with $q \ge 3/2$ because $r_{\rm c}$ rapidly decreases with increase $\theta_{\rm pl, c}$.

\subsubsection{Summary}\label{sec:escapesummary}
%

There exist two possible cases of ${\bm u}_{\rm LC}$ where the radio pulses are not scattered at $r_{\rm LC}$.
First, when ${\bm u}_{\rm LC}$ is significantly inclined with respect to the radio pulses $10^{-1.2} < \theta_{\rm pl, LC} \lesssim 1$ and has the Lorentz factor satisfying $\gamma_{\rm LC} \theta_{\rm pl, LC} (1 + \sigma_{\rm LC})^{1/4} \gtrsim 10^{3.7}$, we obtain $\tau_{\rm LC} < 1$.
In this case, the radio pulses reach the observer without scattering because $\chi(\nu, r)$ decreases rapidly with $r$ for $0 < (a,~b) \lesssim 1.25$ as discussed in Equation (\ref{eq:RadialDependence}).

The second corresponds to the `lack of time' effect, i.e., ${\bm u}_{\rm LC}$ is almost aligned with respect to the radio pulses $\theta_{\rm pl, LC} < 10^{-2.7}$ with $\gamma_{\rm LC} > 10^{2.7}$.
In this case, $r_{\rm in} = \Delta r = r_{\rm c}$, we require $|\tau_{\rm c}| < 1$ when an electron reaches $r_{\rm c}$ and also require $l_{\rm c}(r) > r$ at $r_{\rm LC} < r < r_{\rm c}$.
Using the result $\gamma_{\rm c} > 10^{4.2}$ and $\theta_{\rm pl, c} < 10^{-4.2}$ for $|\tau_{\rm c}| < 1$ ($r_{\rm c} \gtrsim 10^{11.2}~{\rm cm} \approx 10^3 r_{\rm LC}$), $\gamma(r)$ at the range of $r_{\rm LC} < r < 10^{11.2}$ cm should be changed with $r$ as follows (see also Equation (\ref{eq:RadialDepndenceOfCharacteristicRadius}) and Figure \ref{fig:lc_r}).
For the `Narrow' and `Wide' cases, we require that the point `B' ($a < 0.5$) or point `C' ($a \ge 0.5$) in Figure \ref{fig:lc_r} is more distant than $10^{11.2}$ cm.
For example, if $\gamma$ has a constant value ($a = 0$), we require $\gamma > 10^{4.2}$ at $r_{\rm LC}$.
On the other hand, if $a \ge 0.5$ with $\gamma_{\rm LC} > 10^{2.7}$, $\gamma$ should have a terminal value of $\gamma > 10^{4.2}$.
Although the `Inclined' case is a bit complicated, we can constrain the behavior of $\gamma$ by replacing $\gamma$ with $\theta^{-1}_{\rm pl}$ in the above discussion and using the condition $\gamma > \theta^{-1}_{\rm pl}$ ($\Theta_{\rm pl} > 1$) for the `Inclined' case. 
Required values of the exponents $a$ and $b$ change with the value of ${\bm u}_{\rm LC}$, $\sigma_{\rm LC}$ and $\sigma_{\rm c}$.

Lastly, we mention the result obtained by WR78.
Essentially, the `Wide' geometry with scattering at $r_{\rm c} \sim 10^{11.2}$ cm of ours corresponds to the situation which they considered, although their setup is not exactly the same as ours in the radial variations of $\gamma(r)$ and $n_{\rm pl}(r)$.
Our result of $\gamma_{\rm c} \gtrsim 10^{4.8} (1 + \sigma_{\rm c})^{-1/6}$ obtained in Section \ref{sec:escape2} is close to their result of $\gamma > 10^{4.4}$ (see their Equation (16)).
Note that we did not consider the `Wide' case with scattering at $r_{\rm c}$ because $\gamma_{\rm c} < 10^{4.2}$ is also required for the geometry to be `Wide'. 
Also note that they did not account for the constraint at $r_{\rm LC}$, although we require $\gamma_{\rm LC} > 10^{2.7}$ and $\theta_{\rm pl, LC} < 10^{-2.7}$ for $r_{\rm c} > r_{\rm LC}$.

\subsection{Constraints on Pair Multiplicity}\label{sec:Multiplicity}
%
In the last section, we obtain lower limits of $\gamma$ for a given inclination angle $\theta_{\rm pl}$ and a magnetization $\sigma$ of the pulsar wind.
Here, we consider corresponding upper limits of $\kappa$ using Equation (\ref{eq:SpinDownPower}).
Note that the combination of $\kappa \gamma (1 + \sigma) = 10^{10.5}$ is independent of $r$ from energy conservation law and that $\kappa$ alone is also expected to be independent of $r$ from the law of conservation of particle number.
Below, we consider the upper limits of $\kappa$ for the two possible ${\bm u}_{\rm LC}$ of the pulsar wind and we do not consider constraint for the geometry $\Theta_{\rm bm} > \Theta_{\rm pl} > 1$ for simplicity.
\begin{table}[!t]
\caption{
	Lower limits of the Lorentz factor and corresponding upper limits for the pair multiplicity for the two allowed velocities of the pulsar wind at $r_{\rm LC}$ when $r_{\rm e} = 10^7$ cm.
}
\label{tbl:Summary1e7}
\centering
\begin{tabular}{cc}
\hline
$\gamma$       &
$\kappa$       \\
\hline

\multicolumn{2}{c}{Inclined ${\bm u}_{\rm LC}$ ($10^{-1.2} < \theta_{\rm pl, LC} \lesssim 1$)} \\
\hline

$\gamma_{\rm LC} \gtrsim 10^{3.7} \theta^{-1}_{\rm pl, LC} (1 + \sigma_{\rm LC})^{-1/4}$ 
& 
$\kappa \lesssim 10^{6.8} \theta_{\rm pl, LC} (1 + \sigma_{\rm LC})^{-3/4}$
\\

\hline
\hline
\multicolumn{2}{c}{Aligned ${\bm u}_{\rm LC}$ 
($\theta_{\rm pl, LC} < 10^{-2.7}$) and $\theta_{\rm pl, c} < 10^{-4.2}$} \\
\hline

$\gamma_{\rm LC}     > 10^{2.7}$
&
$\kappa \lesssim 10^{7.8} (1 + \sigma_{\rm LC})^{-1}$  
\\

$\gamma_{\rm c}     > 10^{4.2}$      &
$\kappa \lesssim 10^{6.3} (1 + \sigma_{\rm c})^{-1}$  \\

\hline
\end{tabular}
\end{table}

When the pulsar wind is inclined with respect to the radio pulses at $r_{\rm LC}$ ($10^{-1.2} < \theta_{\rm pl, LC} \lesssim 1$), we obtain an upper limit of $\kappa$ by eliminating $\gamma_{\rm LC}$ from $\gamma_{\rm LC} \theta_{\rm pl, LC} (1 + \sigma_{\rm LC})^{1/4} \gtrsim 10^{3.7}$ with the use of Equation (\ref{eq:SpinDownPower}) ($\kappa \gamma(r) (1 + \sigma(r)) = 10^{10.5}$).
We obtain
\begin{equation}\label{eq:Multiplicity1}
\kappa \lesssim 10^{6.8} \theta_{\rm pl, LC} (1 + \sigma_{\rm LC})^{-\frac{3}{4}}.
\end{equation}
The upper limit is $\kappa < 10^{6.8}$ for both $1 + \sigma_{\rm LC} \sim 1$ and $\theta_{\rm pl, LC} \sim 1$.
This upper limit of the pair multiplicity can satisfy $\kappa \gtrsim \kappa_{\rm PWN} = 10^{6.6}$
obtained by Tanaka \& Takahara (2010, 2011) \cite{tt10, tt11}.
However, for $\sigma_{\rm LC} \sim 10^4$, an upper limit becomes $\kappa \lesssim 10^{3.8} \theta_{\rm pl, LC}$ and $\gamma_{\rm LC} \gtrsim 10^{2.7} \theta^{-1}_{\rm pl, LC}$ which can be close to the customarily believed picture of the pulsar wind at the light cylinder \cite{dh82, ha01}.
In other words, $1 + \sigma_{\rm LC} \lesssim 10^{0.2} \theta^{4/3}_{\rm pl, LC}$ is required for $\kappa \ge \kappa_{\rm PWN}$.

For the second case when the pulsar wind is aligned with respect to the radio pulse at $r_{\rm LC}$, we require both $\gamma_{\rm LC} > 10^{2.7}$ ($\theta_{\rm pl, LC} < 10^{-2.7}$) and $\gamma_{\rm c} > 10^{4.2}$ ($\theta_{\rm pl, c} < 10^{-4.2}$).
Using $\kappa \gamma(r) (1 + \sigma(r)) = 10^{10.5}$, we require both
\begin{equation} \label{eq:Multiplicity2}
\kappa \lesssim 10^{7.8} (1 + \sigma_{\rm LC})^{-1}
~{\rm and}~
\kappa \lesssim 10^{6.3} (1 + \sigma_{\rm c} )^{-1}.
\end{equation}
Because $\kappa$ conserves along the flow, $\kappa$ should satisfy both of the two inequalities.
Even for $1 + \sigma_{\rm c} \sim 1$, $\kappa \lesssim 10^{6.3}$ at $r_{\rm c} \sim 10^3 r_{\rm LC}$ is marginal for $\kappa > \kappa_{\rm PWN}$.
For customarily used magnetization $\sigma_{\rm LC} \sim 10^4$, an upper limit is $\kappa \lesssim 10^{3.8} \ll \kappa_{\rm PWN}$.
The results are summarized in Table \ref{tbl:Summary1e7}.
A little bit larger $\kappa$ is allowed for the inclined ${\bm u}_{\rm LC}$ ($\theta_{\rm pl, LC} \sim 1$) than for the aligned ${\bm u}_{\rm LC}$ with respect to the radio pulse beam.

\subsection{Dependence on the Size of Emission Region}\label{sec:EmissionRegion}
%
\begin{figure}[t]
\begin{minipage}{0.5\hsize}
\begin{center}
\includegraphics[scale=0.3]{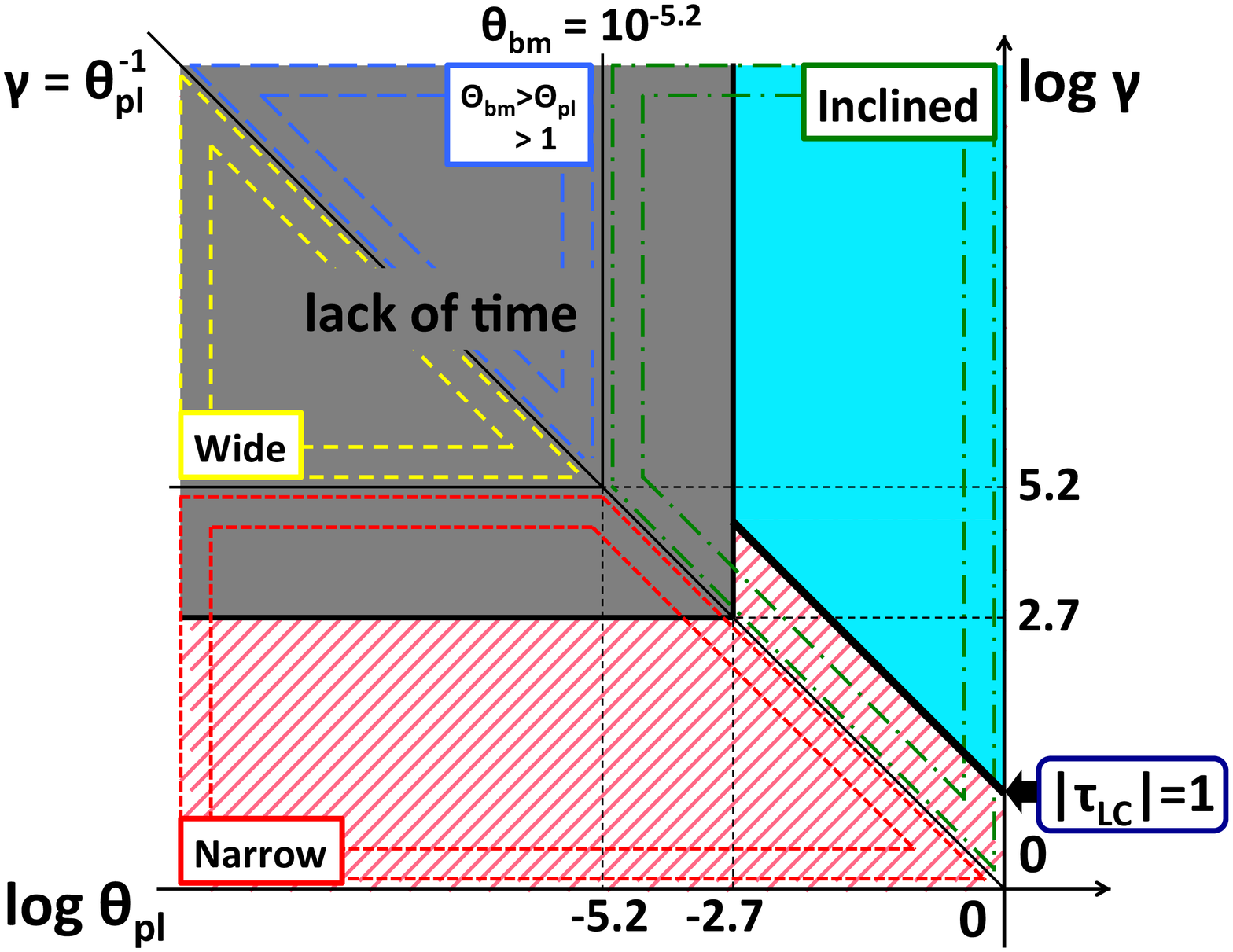}
\end{center}
\end{minipage}
\begin{minipage}{0.5\hsize}
\begin{center}
\includegraphics[scale=0.3]{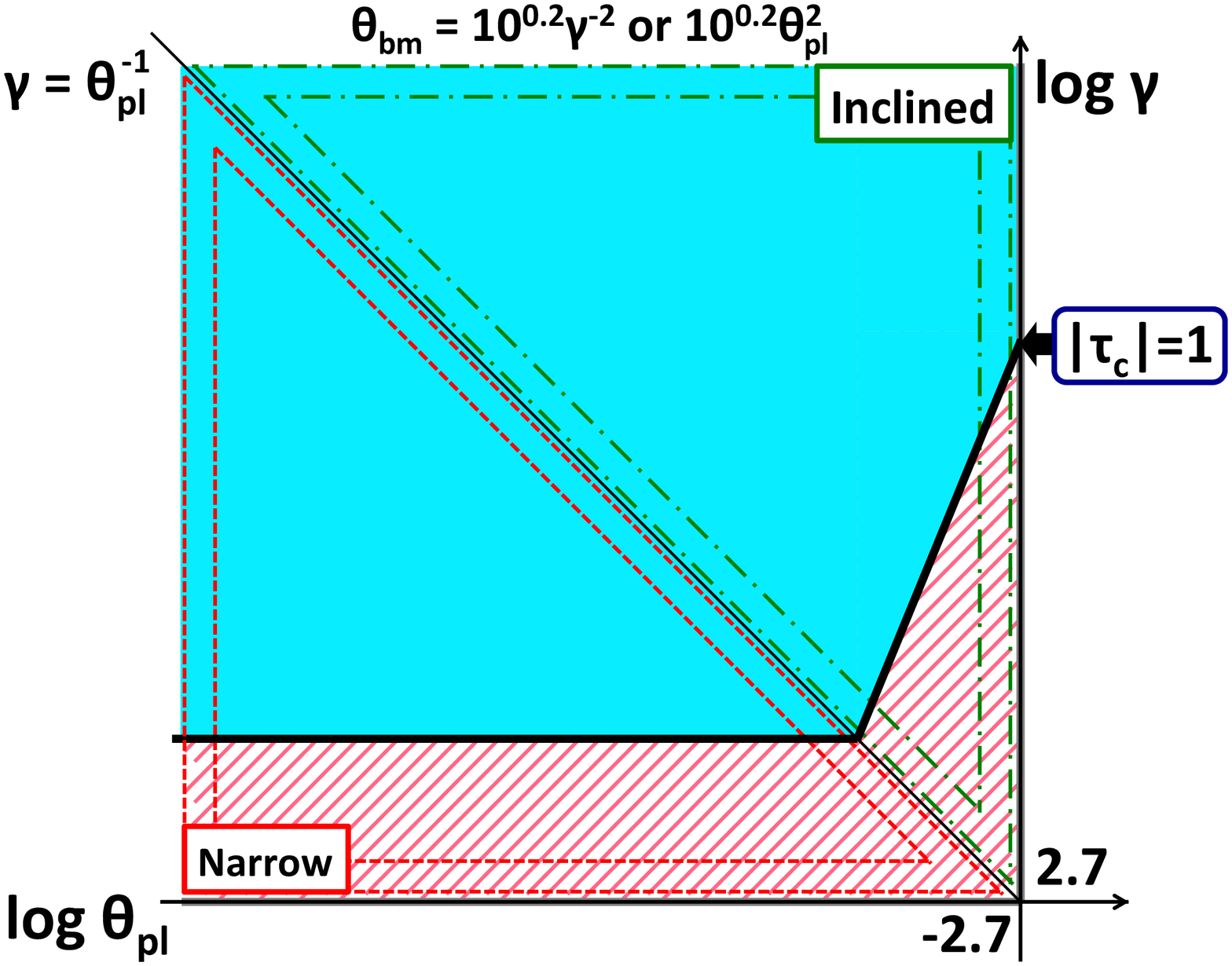}
\end{center}
\end{minipage}
\caption{
	The $\gamma - \theta_{\rm pl}$ diagrams at $r_{\rm LC}$ (left) and $r_{\rm c}$ (right).
	We take different emission region size of $r_{\rm e} = 10^3$ cm from Figures \ref{fig:ConstraintAtRlc1e7} and \ref{fig:ConstraintAtRc1e7} (see also Tables \ref{tbl:1e7rlc} and \ref{tbl:1e7rc}).
	The `lack of time' region (gray in color) on the left panel is the same extent as Figures \ref{fig:ConstraintAtRlc1e7}.
	The shaded region (pink in color) is forbidden region for both panels.
	The 'Narrow' and `Inclined' areas expand compared with Figures \ref{fig:ConstraintAtRlc1e7} and \ref{fig:ConstraintAtRc1e7} because $\theta_{\rm bm} \propto r_{\rm e}$ in Equation (\ref{eq:ThetaBeam}). 
	For the left panel, three lines $\gamma = \theta^{-1}_{\rm pl}$, $\gamma = 10^{5.2}$ and $\theta_{\rm pl} = 10^{-5.2}$ divides scattering geometries, while we do not find the `Wide' and $\Theta_{\rm bm} > \Theta_{\rm pl} > 1$ areas for the right panel.
	$|\tau| = 1$ lines are also different from and $|\tau| < 1$ region becomes wider than Figures \ref{fig:ConstraintAtRlc1e7} and \ref{fig:ConstraintAtRc1e7}.
	The $|\tau_{\rm LC}| = 1$ line in the `Inclined' region on the left panel corresponds to $\gamma_{\rm LC} \gtrsim 10^{1.7} \theta^{-1}_{\rm pl, LC} (1 + \sigma_{\rm LC})^{-1/4}$.
	We adopt $1 < 1 + \sigma_{\rm LC} \le 10^4$ in the figure, i.e., y-intercept of the $|\tau_{\rm LC}| = 1$ line on the left panel is $10^{0.7} \le \gamma < 10^{1.7}$, for example.
	The $|\tau_{\rm c}| = 1$ lines on the right panel correspond to	$\gamma_{\rm c} \gtrsim 10^{3.4} (1 + \sigma_{\rm c})^{-1/10}$ for the `Narrow' area and $\gamma_{\rm c} \gtrsim 10^{3.4} \Theta^{3/5}_{\rm pl, c} (1 + \sigma_{\rm c})^{-1/10}$ for the `Inclined' area.
}
\label{fig:ConstraintAt1e3}
\end{figure}

We assume $r_{\rm e} = 10^7$ cm in the above calculations.
Here, we discuss the constraints on $\gamma$ and $\kappa$ assuming Equation (\ref{eq:ThetaBeam}) with $r_{\rm e} = 10^3$ cm for example.
The dependence on $r_{\rm e}$ ($10^3 \le r_{\rm e} \le 10^7$ cm) is described explicitly in Tables \ref{tbl:1e7rlc} and \ref{tbl:1e7rc}.
When we take a different value of $r_{\rm e}$, the brightness temperature $T_{\rm b}$ (Equation (\ref{eq:BrightnessTemperature})) and the integrals $I_{\rm Narrow}$ and $I_{\rm Inclined}$ (Equations (\ref{eq:NarrowBeam}) and (\ref{eq:Inclined})) are changed.
In Tables \ref{tbl:1e7rlc} and \ref{tbl:1e7rc}, we find that the optical depth for the `Narrow' and `Inclined' cases is proportional to $r^2_{\rm e}$.
This is because $I_{\rm Narrow}$ and $I_{\rm Inclined}$ are proportional to $r^4_{\rm e}$ and $T_{\rm b}$ is proportional to $r^{-2}_{\rm e}$.
On the other hand, for the `Wide' case, the optical depth is proportional to $r^{-2}_{\rm e}$ because $I_{\rm Wide}(\nu) \sim -1$ whose value does not depend on $\theta_{\rm bm}$ in the range of $\nu_0 \lesssim \nu < \Theta^2_{\rm bm} \nu_0$.
Note that the layout of scattering geometry on the $\gamma - \theta_{\rm pl}$ diagrams (Figure \ref{fig:ConstraintAt1e3}) is also changed where the `Narrow' and `Inclined' areas spread on the planes compared with those in Figures \ref{fig:ConstraintAtRlc1e7} and \ref{fig:ConstraintAtRc1e7}.

We obtain the lower limits of $\gamma$ and the upper limits of $\kappa$ in the same manner as the case of $r_{\rm e} = 10^7$ cm.
Figure \ref{fig:ConstraintAt1e3} shows the resultant $\gamma - \theta_{\rm pl}$ diagrams both at $r_{\rm LC}$ (left) and $r_{\rm c}$ (right).
Obtained lower limits of $\gamma$ and upper limits of $\kappa$ are summarized in Table \ref{tbl:Summary1e3}.

\begin{table}[!t]
\caption{
Lower limits of the Lorentz factor and corresponding upper limits for the pair multiplicity for the two possible structures of the pulsar wind at $r_{\rm LC}$ when $10^3 \le r_{\rm e} \le 10^7$ cm.
}
\label{tbl:Summary1e3}
\centering
\begin{tabular}{cc}
\hline
$\gamma$       &
$\kappa$ \\
\hline

\multicolumn{2}{c}{Inclined ${\bm u}_{\rm LC}$ 
($\max(10^{-2.7}, 10^{-5.2} r_{\rm e, 3}) < \theta_{\rm pl, LC} \lesssim 1$)
} \\
\hline

$\gamma_{\rm LC} \gtrsim 10^{1.7} r^{1/2}_{\rm e, 3} \theta^{-1}_{\rm pl, LC} (1 + \sigma_{\rm LC})^{-1/4}$ 
&
$\kappa \lesssim 10^{8.8} r^{-1/2}_{\rm e, 3} \theta_{\rm pl, LC} (1 + \sigma_{\rm LC})^{-3/4}$ 
\\

\hline
\hline
\multicolumn{2}{c}{Aligned ${\bm u}_{\rm LC}$ 
($\theta_{\rm pl, LC} < 10^{-2.7}$)
and $\theta_{\rm pl, c} < \gamma^{-1}_{\rm c}$} \\
\hline

$\gamma_{\rm LC} \gtrsim 10^{2.7}$ 
&
$\kappa \lesssim 10^{7.8} (1 + \sigma_{\rm LC})^{-1}$  
\\

$\gamma_{\rm c} \gtrsim 10^{3.4} r^{1/5}_{\rm e, 3}  (1 + \sigma_{\rm c})^{-1/10}$ &
$\kappa \lesssim 10^{7.1} r^{-1/5}_{\rm e, 3} (1 + \sigma_{\rm c})^{-9/10}$ \\
%
%


%
\hline
\hline
\multicolumn{2}{c}{Aligned ${\bm u}_{\rm LC}$ 
($\theta_{\rm pl, LC} < 10^{-2.7}$)
and $\theta_{\rm pl, c} > \gamma^{-1}_{\rm c}$} \\
\hline

$\gamma_{\rm LC} \gtrsim 10^{2.7}$ 
&
$\kappa \lesssim 10^{7.8} (1 + \sigma_{\rm LC})^{-1}$  
\\

$\gamma_{\rm c} \gtrsim 10^{3.4} r^{1/5}_{\rm e, 3} \Theta^{3/5}_{\rm pl, c} (1 + \sigma_{\rm c})^{-1/10}$ &
$\kappa \lesssim 10^{7.1} r^{-1/5}_{\rm e, 3} \Theta^{-3/5}_{\rm pl, c} (1 + \sigma_{\rm c})^{-9/10}$ \\
\hline
\end{tabular}
\end{table}

At $r_{\rm LC}$ ($\theta_{\rm bm}(r_{\rm LC}) \approx 10^{-5.2}$), we find two allowed regions on the diagram in the left panel of Figure \ref{fig:ConstraintAt1e3}.
First is when the pulsar wind has a significant non-radial motion $10^{-2.7} < \theta_{\rm pl, LC} \lesssim 1$.
We require $\gamma_{\rm LC} \theta_{\rm pl, LC} (1 + \sigma_{\rm LC})^{1/4} \gtrsim 10^{1.7} r^{1/2}_{\rm e, 3}$ for $|\tau_{\rm LC}| < 1$ and no scattering occurs beyond $r_{\rm LC}$ for the moderate values of the exponents $a$ and $b$.
We also find that the non-relativistic pulsar wind $\beta_{\rm LC} \ll 1$ is unfavorable even for such a small opening angle of the radio beam $\theta_{\rm bm, LC} = 10^{-5.2}$ with $1 + \sigma_{\rm LC} \approx 10^4$.

Secondly, the region which satisfies $\gamma_{\rm LC} > 10^{2.7}$ and $\theta_{\rm pl, LC} < 10^{-2.7}$ is also allowed to escape from scattering at $r_{\rm LC}$ due to the `lack of time' effect.
In this case, in addition, we require $|\tau_{\rm c}| < 1$ at $r_{\rm c}$ ($> r_{\rm LC}$).
The right panel of Figure \ref{fig:ConstraintAt1e3} shows the $\gamma - \theta_{\rm pl}$ diagram at $r_{\rm c}$.
We do not find the `Wide' and $\Theta_{\rm bm} > \Theta_{\rm pl} > 1$ geometries on the diagram because $\theta_{\rm bm}(r_{\rm c})$ for $r_{\rm e} = 10^3$ cm is much smaller than that for $r_{\rm e} = 10^7$ cm.
The region which satisfies $|\tau_{\rm c}| < 1$ is $\gamma_{\rm c} \gtrsim 10^{3.4} r^{1/5}_{\rm e, 3} (1 + \sigma_{\rm c})^{-1/10}$ for the `Narrow' case and $\gamma_{\rm c} \gtrsim 10^{3.4} r^{1/5}_{\rm e, 3} \Theta^{3 / 5}_{\rm pl, c} (1 + \sigma_{\rm c})^{-1/10}$ for the `Inclined' case.
Corresponding $r_{\rm c}$ is larger than $10^{9.6} r^{2/5}_{\rm e, 3}$ cm $= 10^{1.4} r^{2/5}_{\rm e, 3} r_{\rm LC}$.
It is important to note that the constraint at $r_{\rm c}$ very weakly depends on $r_{\rm e}$ as $r^{1/5}_{\rm e}$.

Accordingly, we obtain upper limits of $\kappa$ with the help of Equation (\ref{eq:SpinDownPower}).
When the pulsar wind is inclined with respect to the radio pulse at $r_{\rm LC}$ ($10^{-2.7} < \theta_{\rm pl, LC} \lesssim 1$), we obtain
\begin{equation}\label{eq:Multiplicity3}
	\kappa \lesssim 10^{8.8} r^{-\frac{1}{2}}_{\rm e, 3} \theta_{\rm pl, LC} (1 + \sigma_{\rm LC})^{-\frac{3}{4}}.
\end{equation}
We require $\sigma_{\rm LC} \lesssim 10^3 \ll 10^4$ for $\kappa > \kappa_{\rm PWN}$.
When the pulsar wind is aligned with respect to the radio pulse at $r_{\rm LC}$ ($\theta_{\rm pl, LC} < 10^{-2.7}$ and $\gamma_{\rm LC} > 10^{2.7}$), we obtain
\begin{equation}\label{eq:Multiplicity4}
	\kappa \lesssim 10^{7.8} (1 + \sigma_{\rm LC})^{-1}
	~{\rm and}~  \left\{
	\begin{array}{ll}
	\kappa \lesssim 10^{7.1} r^{-\frac{1}{5}}_{\rm e, 3} (1 + \sigma_{\rm c})^{-\frac{9}{10}} & \mbox{ for `Narrow',} \\
	\kappa \lesssim 10^{7.1} r^{-\frac{1}{5}}_{\rm e, 3} \Theta^{-\frac{3}{5}}_{\rm pl, c} (1 + \sigma_{\rm c})^{-\frac{9}{10}} & \mbox{ for `Inclined'.}
	\end{array} 
	\right.
\end{equation}
$\kappa > \kappa_{\rm PWN}$ is attainable for both the 'Narrow' and 'Inclined' cases again.

We obtain the lower limits of $\gamma$ and the upper limits of $\kappa$ for different sizes of the emission region $r_{\rm e}$.
Basically, as is found from Table \ref{tbl:Summary1e3}, the smaller the emission region size becomes, the easier the radio pulses escape from scattering, i.e., small $\gamma$ and large $\kappa$ are allowed.
We obtain the most optimistic constraint for large $\kappa$ ($\kappa \lesssim 10^{8.8}$ at the uppermost row of Table \ref{tbl:Summary1e3}), when $\theta_{\rm pl, LC} \sim 1$ (inclined ${\bm u}_{\rm LC}$), $1 + \sigma_{\rm LC} \sim 1$ and $r_{\rm e} = 10^3$ cm.
Combined with $\kappa \gtrsim \kappa_{\rm PWN} = 10^{6.6}$, we can write the pulsar wind properties as $10^{1.7} \lesssim \gamma \lesssim 10^{3.9}$ and $\kappa_{\rm PWN} \lesssim \kappa \lesssim 10^{8.8}$.
Although all these constraints are at $r_{\rm LC}$, the radio pulse can escape from scattering and $\kappa \gtrsim \kappa_{\rm PWN}$ is satisfied beyond $r_{\rm LC}$ because $\gamma(r)(1 + \sigma(r)) \approx \gamma(r) =$ constant beyond $r_{\rm LC}$ for $1 + \sigma_{\rm LC} \sim 1$ from Equation (\ref{eq:SpinDownPower}) and conservation of particle number ($\kappa =$ constant).
Note that we obtain $10^{1.2} \lesssim \gamma_{\rm LC} \lesssim 10^{1.9}$ and $\kappa_{\rm PWN} \lesssim \kappa \lesssim 10^{7.3}$ for $1 + \sigma_{\rm LC} \sim 10^2$, and we require $\gamma(r) (1 + \sigma(r)) =$ constant and also $\kappa =$ constant beyond $r_{\rm LC}$.

\section{Summary}\label{sec:Summary}
%

To constrain the pulsar wind properties, we study induced Compton scattering by a relativistically moving cold plasma.
Induced Compton scattering is $\theta^4_{\rm bm} k_{\rm B} T_{\rm b}(\nu) / m_{\rm e} c^2$ times significant compared with spontaneous scattering for the non-relativistic case.
However, for scattering by the relativistically moving plasma, scattering geometry of the system changes the scattering coefficient significantly.
We consider fairly general geometries of scattering in the observer frame and obtain the scattering coefficient for induced Compton scattering off the photon beam.
On the other hand, we do not take into account the magnetic field effects and the scattering off the background photons in this paper.

We obtain approximate expressions of the scattering coefficient for three geometries corresponding to the `Narrow' ($1 > \Theta^2_{\rm bm} + \Theta^2_{\rm pl}$), `Inclined' ($\Theta^2_{\rm pl} > 1 + \Theta^2_{\rm bm}$) and `Wide' ($\Theta_{\rm bm} > 1 > \Theta_{\rm pl}$) cases, while the scattering coefficient for $\Theta_{\rm bm} > \Theta_{\rm pl} > 1$ is obtained numerically.
Behavior of the scattering coefficient against a given scattering geometry is governed by a simple combination of four factors.
In addition to the solid angle factor $\theta^4_{\rm bm}$ appearing even for the non-relativistic case, there exist three relativistic effects; the factor independent of scattering geometry $\gamma^{-3}$ and the other two factors depending on geometry, the aberration factor $D^{-2}_1$ and the frequency shift factor $D / D_1$.
When the photon beam is inside the $\gamma^{-1}$ cone of the plasma beam (the `Narrow' case), the aberration factor increases the scattering coefficient by a factor of $\sim \gamma^4$ (up to $\gamma \theta_{\rm bm} \sim 1$).
On the other hand, when the plasma velocity is significantly inclined with respect to the photon beam (the `Inclined' case), this factor of $\gamma^4$ does not appear.
The frequency shift factor is important when the photon beam is wider than the $\gamma^{-1}$ cone of the plasma beam (the `Wide' case) and is rather complex and mostly increases the absolute value of the scattering coefficient compared with the non-relativistic case.
Basically, the `Inclined' case gives the smallest and the 'Wide' case gives the largest scattering coefficient, i.e., the $\Theta_{\rm bm} > \Theta_{\rm pl} > 1$ case is in between.

We apply induced Compton scattering to the Crab pulsar, where the high $T_{\rm b}(\nu)$ radio pulses go through the relativistic pulsar wind and constrain the pulsar wind properties by imposing the condition of the optical depth being smaller than unity.
We introduce the characteristic scattering radius $r_{\rm c}$ where the `lack of time' effect prevents scattering at $r < r_{\rm c}$.
We evaluate the scattering optical depth for both $r_{\rm in } = r_{\rm LC}$ and $r_{\rm in } = r_{\rm c}$ cases.
We consider more general scattering geometries than WR78 and also study the dependence on the size of the emission region $10^3 \le r_{\rm e} \le 10^7$ cm which directly affects the opening angle of the radio pulses $\theta_{\rm bm}(r)$.
Allowable pulsar wind velocities at $r_{\rm LC}$ (${\bm u}_{\rm LC}$) and at $r_{\rm c}$ (${\bm u}_{\rm c}$) are explored assuming the canonical value of the magnetization $1 < 1 + \sigma \lesssim 10^4$.

The two pulsar wind velocities ${\bm u}_{\rm LC}$ are allowed for radio pulses to escape from scattering at $r_{\rm LC}$.
One is that the plasma velocity is inclined with respect to the photon beam ($\theta_{\rm pl, LC} \sim 1$).
When $\gamma_{\rm LC} \gtrsim 10^{1.7} r^{1/2}_{\rm e, 3} \theta^{-1}_{\rm pl, LC} (1 + \sigma_{\rm LC})^{-1/4}$ is satisfied, the radio pulses reach the observer without scattering for moderate radial variation of $\gamma(r)$ and $\theta_{\rm pl}(r)$ where $\gamma \propto r^a$ and $\theta_{\rm pl} \propto r^{-b}$ with $0 < (a,~b) \lesssim 1.25$.
The other is when the plasma velocity is aligned with respect to the photon beam ($\theta_{\rm pl, LC} < 10^{-2.7}$).
We require the lower limit $\gamma_{\rm LC} \gtrsim 10^{2.7}$ for the `lack of time' effect preventing scattering at $r_{\rm LC}$.
In this case, we also require the optical depth at $r_{\rm c} \gtrsim 10^{9.6} r^{2/5}_{\rm e, 3}$ cm $= 10^{1.4} r^{2/5}_{\rm e, 3} r_{\rm LC}$ to be less than unity, where $r_{\rm c}$ ($= l_{\rm c}$) depends on $\gamma_{\rm c}$ or $\theta_{\rm c}$ (Equation (\ref{eq:CharacteristicRadius})).
For example, we require $\gamma_{\rm c} \gtrsim 10^{3.4} r^{1/5}_{\rm e, 3} (1 + \sigma_{\rm c})^{-1/10}$ for the completely aligned case $\theta_{\rm pl} = 0$.
Basically, the smaller the emission region size and the larger the inclination angle of the pulsar wind become, the smaller $\gamma$ is allowed.

We discussed upper limits of the pair multiplicity using obtained constraints on the velocities of the Crab pulsar wind and Equation (\ref{eq:SpinDownPower}).
In principle, $\kappa \gtrsim \kappa_{\rm PWN} \equiv 10^{6.6}$ \cite[][]{tt10, tt11} is possible although we require $1 + \sigma_{\rm LC} \ll 10^4$, i.e., customarily used value $1 + \sigma_{\rm LC} \approx 10^4$ contradicts $\kappa > \kappa_{\rm PWN}$.
The most optimistic constraint which allows large $\kappa$ is obtained when $\theta_{\rm pl, LC} \sim 1$ and $r_{\rm e} = 10^3$ cm (Equation (\ref{eq:Multiplicity3})).
In this case with $\kappa \gtrsim \kappa_{\rm PWN}$, we can write the pulsar wind properties as $10^{1.7} \lesssim \gamma \lesssim 10^{3.9}$ and $\kappa_{\rm PWN} \lesssim \kappa \lesssim 10^{8.8}$ for $1 + \sigma_{\rm LC} \sim 1$ and $10^{1.2} \lesssim \gamma \lesssim 10^{1.9}$ and $\kappa_{\rm PWN} \lesssim \kappa \lesssim 10^{7.3}$ for $1 + \sigma_{\rm LC} \sim 10^2$.
Note that all these constraints are at $r_{\rm LC}$ and we also require moderate radial variation of $\theta_{\rm pl}(r)$ and $\gamma(r)$ ($\propto (1 + \sigma(r))^{-1}$) beyond $r_{\rm LC}$.

\section*{Acknowledgment}
S. J. T. would like to thank Y. Ohira, R. Yamazaki, T. Inoue and S. Kisaka for useful discussion.
We would also like to thank the anonymous referees for a meticulous reading of the manuscript and very helpful comments.
This work is supported by JSPS Research Fellowships for Young Scientists (S.J.T. 2510447).

\appendix

\section{Numerical Integration}\label{app:NumericalIntegration}
\begin{figure}[t]
\begin{minipage}{0.5\hsize}
\begin{center}
\includegraphics[scale=0.6]{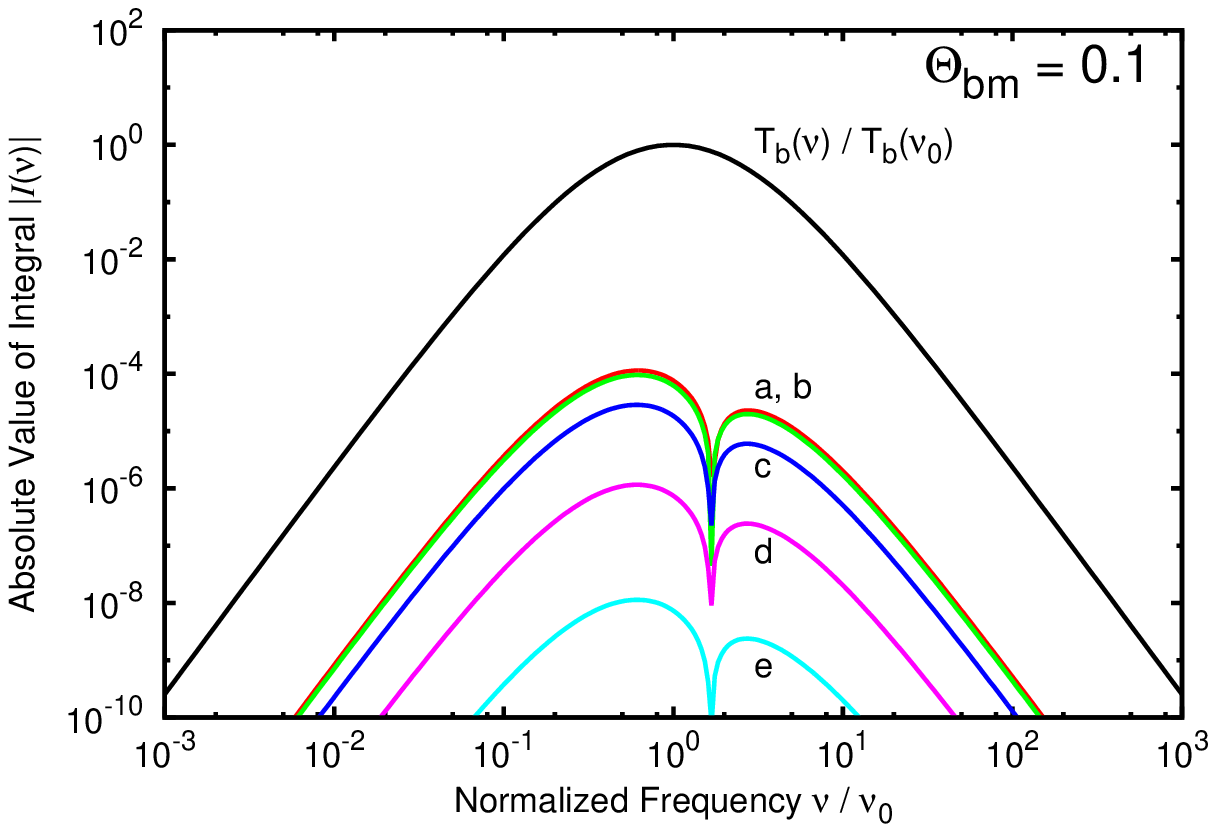}
\end{center}
\end{minipage}
\begin{minipage}{0.5\hsize}
\begin{center}
\includegraphics[scale=0.6]{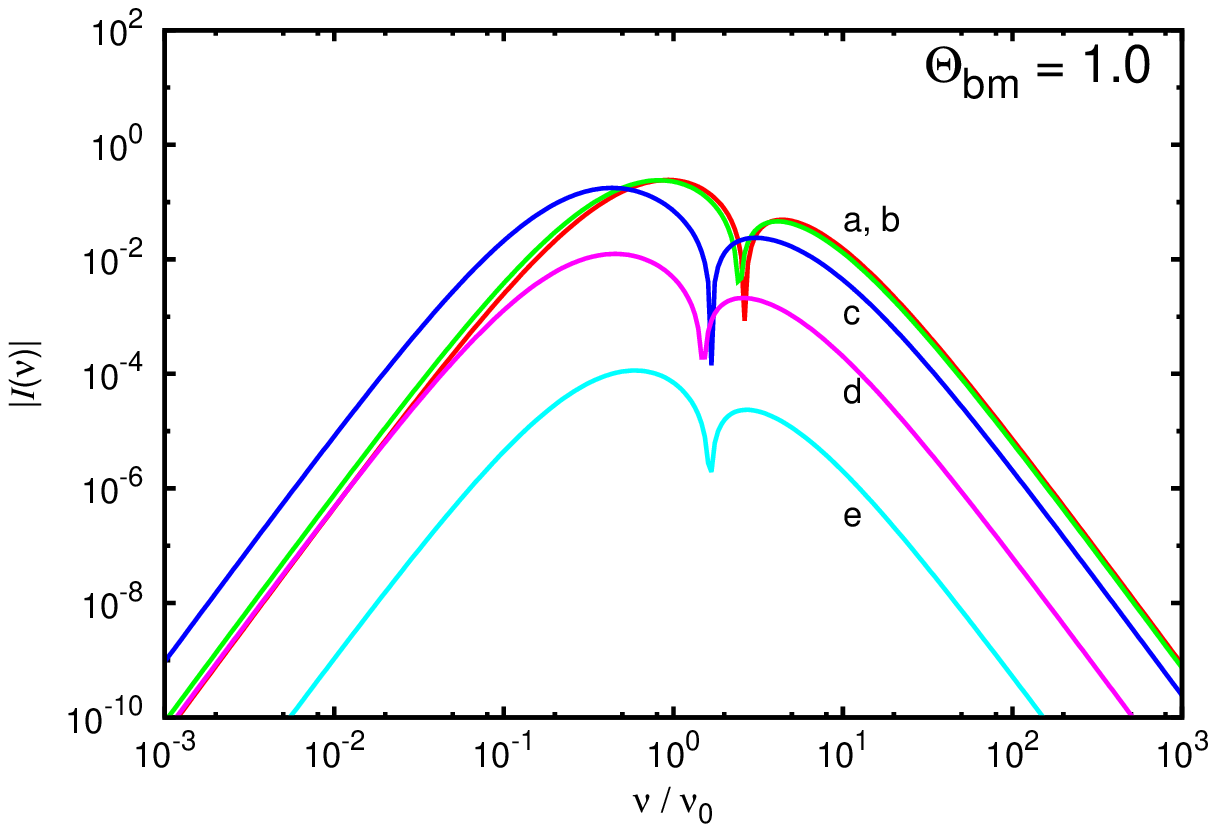}
\end{center}
\end{minipage}
\begin{minipage}{0.5\hsize}
\begin{center}
\includegraphics[scale=0.6]{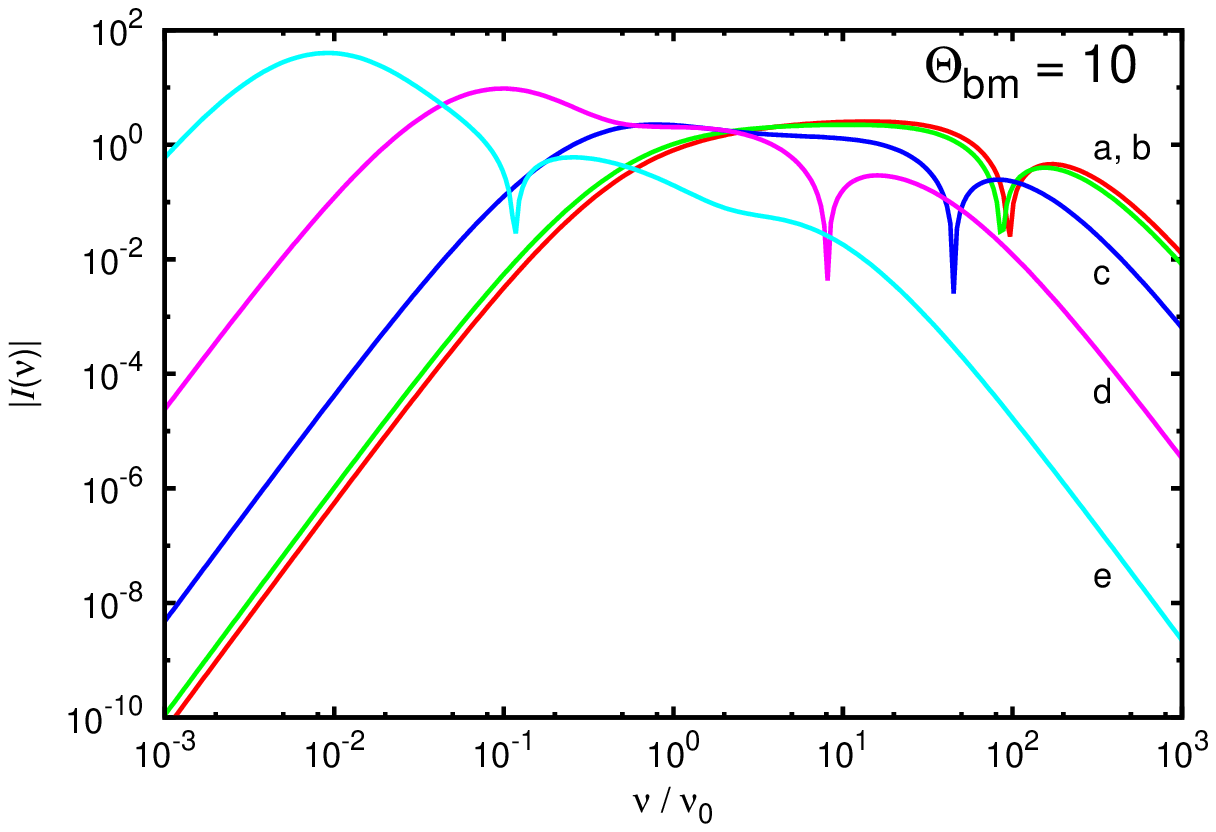}
\end{center}
\end{minipage}
\begin{minipage}{0.5\hsize}
\begin{center}
\includegraphics[scale=0.3]{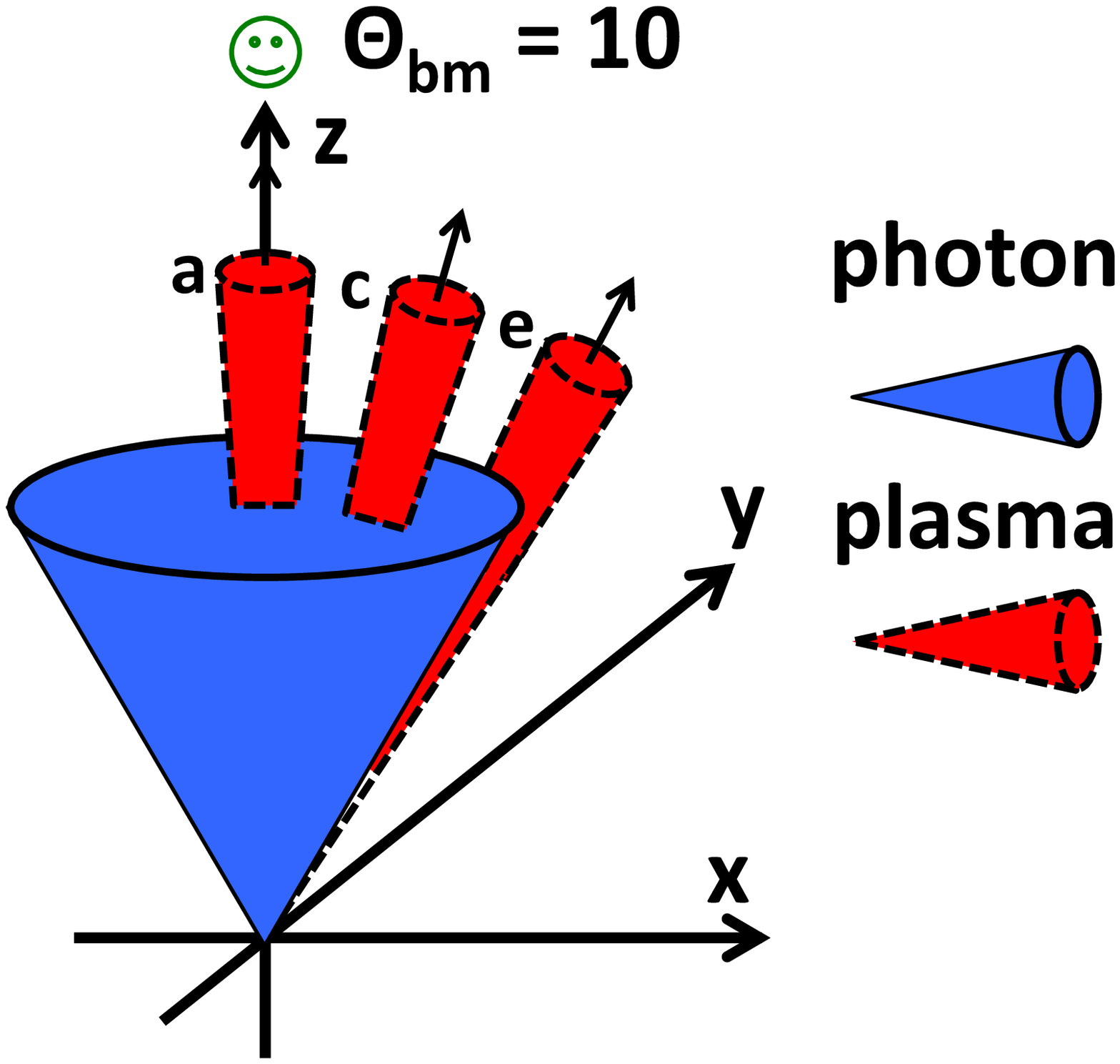}
\end{center}
\end{minipage}
\caption{
Plots of the integral $I(\nu, \theta_{\rm bm}, \theta_{\rm pl}, \gamma)$ and the sketch of scattering geometry (bottom-right).
To see the dependence on $\Theta_{\rm pl}$, we fix $\Theta_{\rm bm}$ for each panel, where top-left: $\Theta_{\rm bm} = 10^{-1}$, top-right: $\Theta_{\rm bm} = 1$ and bottom-left: $\Theta_{\rm bm} = 10$, respectively.
Each line is for a different value of $\Theta_{\rm pl}$, where `line a': $\Theta_{\rm pl} = 0$, `line b': $= 0.3$, `line c': $= 1$, `line d': $= 3$, and `line e': $= 10$, respectively.
We set $\gamma = 10^2$, $p_1 = 3$ and $p_2 = -5$.
}
\label{fig:ChangeThetaPlasma}
\end{figure}

\begin{figure}[t]
\begin{minipage}{0.5\hsize}
\begin{center}
\includegraphics[scale=0.6]{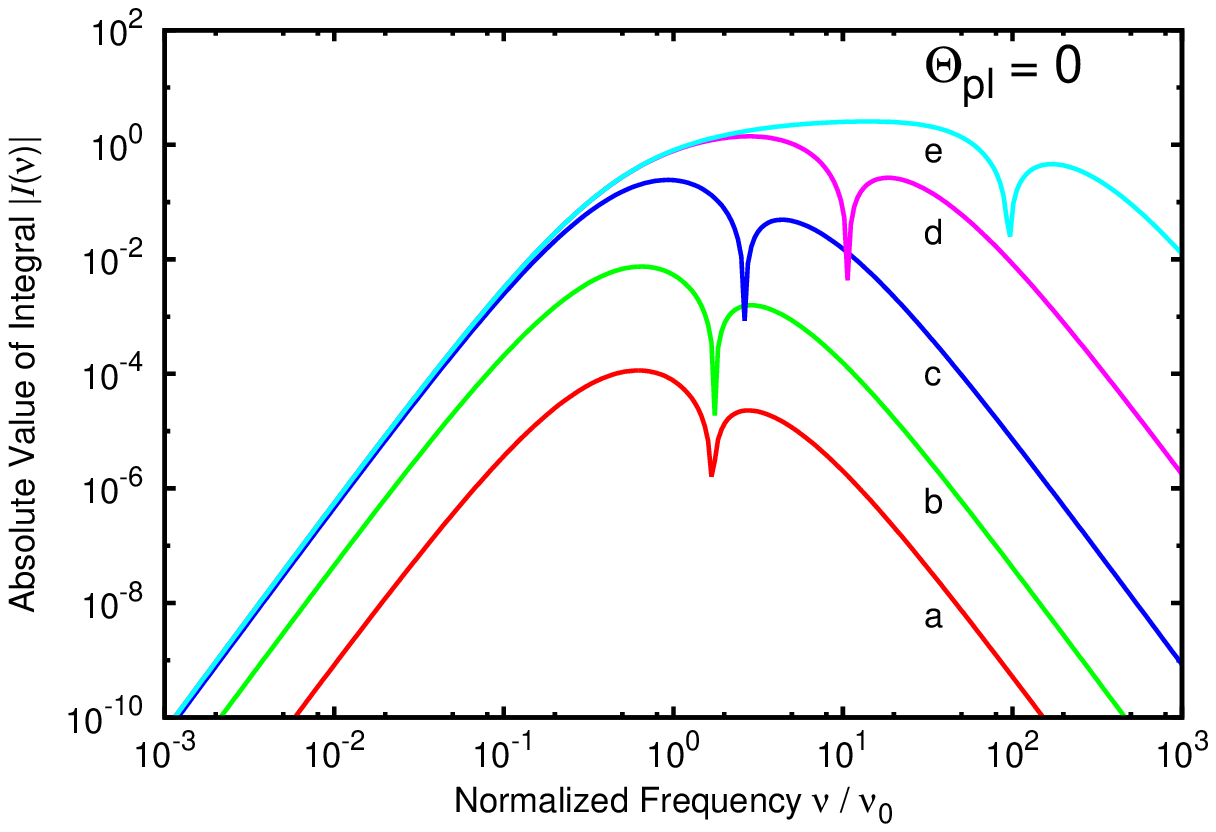}
\end{center}
\end{minipage}
\begin{minipage}{0.5\hsize}
\begin{center}
\includegraphics[scale=0.6]{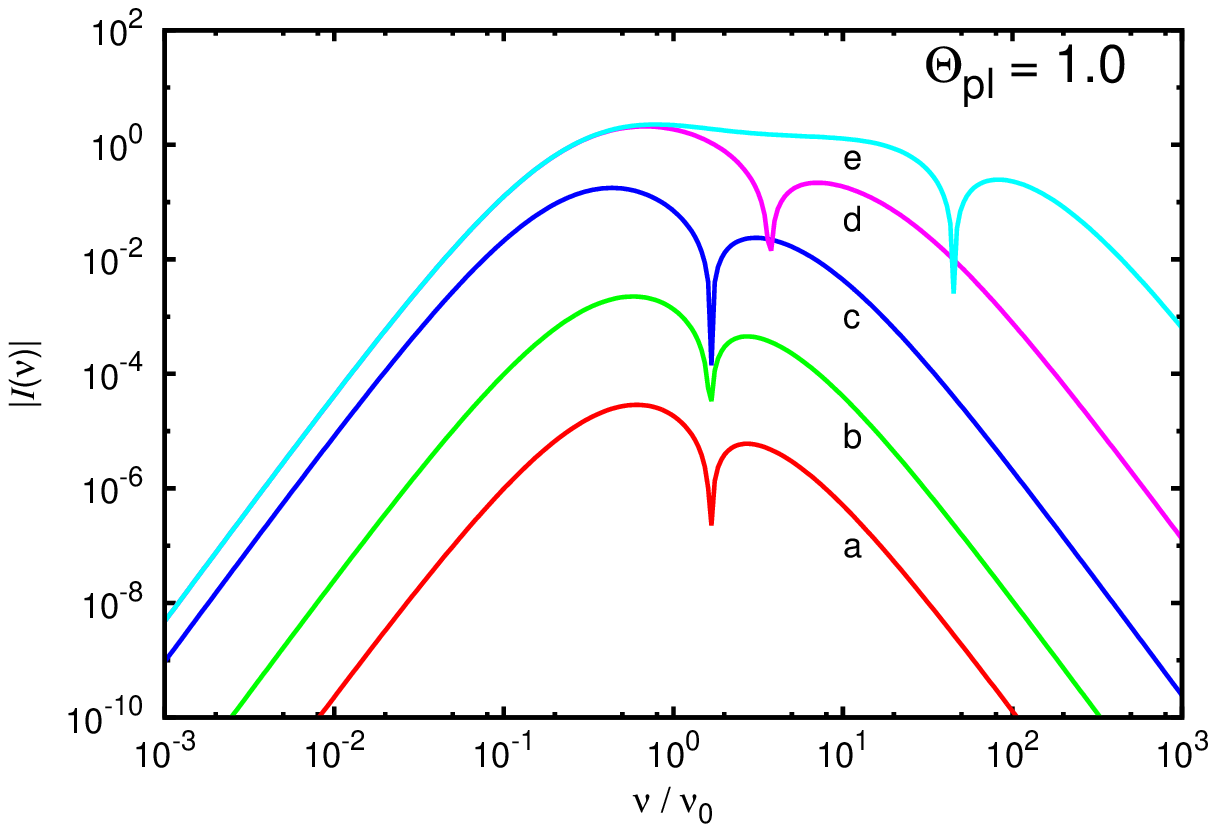}
\end{center}
\end{minipage}
\begin{minipage}{0.5\hsize}
\begin{center}
\includegraphics[scale=0.6]{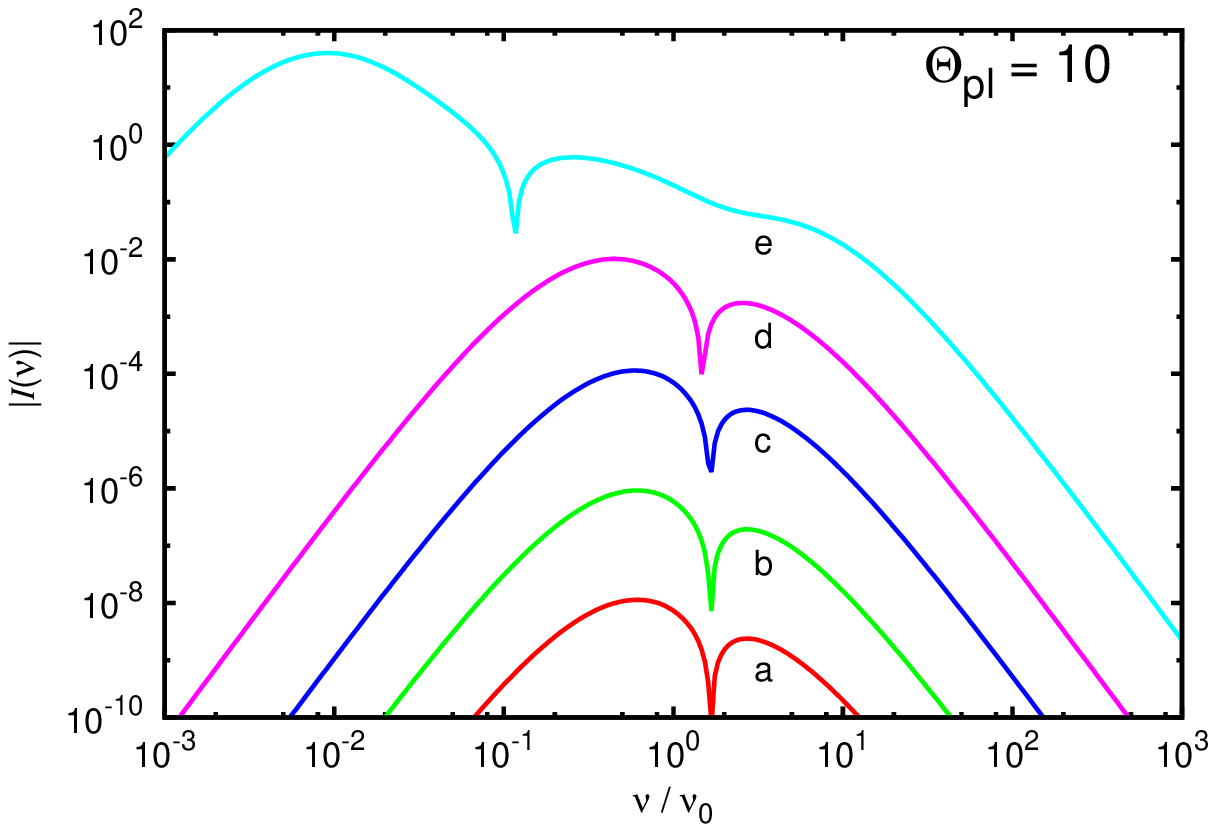}
\end{center}
\end{minipage}
\begin{minipage}{0.5\hsize}
\begin{center}
\includegraphics[scale=0.3]{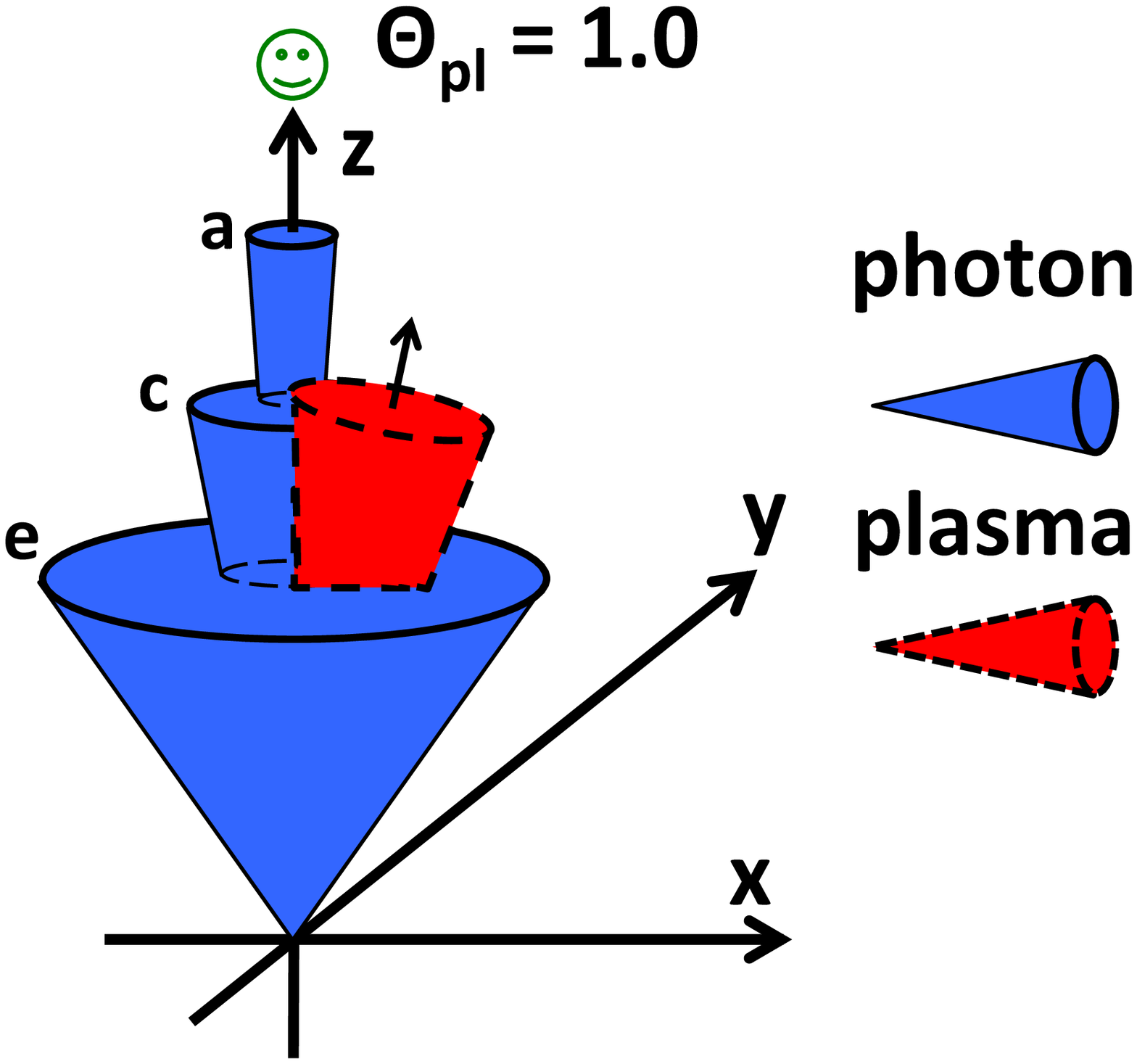}
\end{center}
\end{minipage}
\caption{
Plots of the integral $I(\nu, \theta_{\rm bm}, \theta_{\rm pl}, \gamma)$ and the sketch of scattering geometry (bottom-right).
To see the dependence on $\Theta_{\rm bm}$, we fix $\Theta_{\rm pl}$ for each panel, where top-left: $\Theta_{\rm pl} = 0$, top-right: $\Theta_{\rm pl} = 1$ and bottom-left: $\Theta_{\rm pl} = 10$, respectively.
Each line is for a different value of $\Theta_{\rm bm}$, where `line a': $\Theta_{\rm bm} = 0.1$, `line b': $= 0.3$, `line c': $= 1$, `line d': $= 3$, and `line e': $= 10$, respectively.
We set $\gamma = 10^2$, $p_1 = 3$ and $p_2 = -5$.
}
\label{fig:ChangeThetaPhoton}
\end{figure}

%
We show results of numerical integration of $I(\nu, \gamma, \theta_{\rm bm}, \theta_{\rm pl})$ (Equation (\ref{eq:DefineIntegral})).
We focus on the situation $0 \le (\theta_{\rm pl},~\theta_{\rm bm}) \lesssim 1$ and $\gamma \gg 1$, and then the integral $I(\nu)$ depends on the normalized angles $\Theta_{\rm bm} \equiv \gamma \theta_{\rm bm}$ and $\Theta_{\rm pl} \equiv \gamma \theta_{\rm pl}$ rather than on $\theta_{\rm bm}$, $\theta_{\rm pl}$ and $\gamma$, separately.
As seen in Section \ref{sec:AnalyticEstimates}, the behavior of $I(\nu)$ is very different for the value of $\Theta_{\rm bm}$ and $\Theta_{\rm pl}$, i.e., different scattering geometries.
To obtain the results of Figures \ref{fig:ChangeThetaPlasma} and \ref{fig:ChangeThetaPhoton}, we set $\gamma = 10^2$ and adopt the broken power-law spectrum with $p_1 = 3$ and $p_2 = -5$ (Equation (\ref{eq:PhotonFrequencyDistribution})).
The figures show absolute values of the integral $I(\nu)$ versus frequency $\nu$ for different sets of parameters $\Theta_{\rm bm}$ and $\Theta_{\rm pl}$.
All the lines in these figures have a discontinuity where the sign of the integral $I(\nu)$ changes.
The sign of the integral $I(\nu)$ is positive at high frequency side where the photon number decreases and vice versa.

Before describing details of Figures \ref{fig:ChangeThetaPlasma} and \ref{fig:ChangeThetaPhoton}, we mention that the approximated forms studied in Section \ref{sec:AnalyticEstimates} can describe behaviors of most of lines in the figures.
Behaviors of lines with no frequency shift is described by $I_{\rm Narrow}$ and $I_{\rm Inclined}$ and behaviors of lines whose discontinuity point shifted to $\nu > \nu_0$ is described by $I_{\rm Wide}$.
Only behaviors of `line d' and `line e' in the bottom-left panel in Figure \ref{fig:ChangeThetaPlasma} and of `line e' in the bottom-left panel in Figure \ref{fig:ChangeThetaPhoton} are not explained by these three approximated forms corresponding to $\Theta_{\rm bm} > \Theta_{\rm pl} > 1$ which we will discuss later.

Figure \ref{fig:ChangeThetaPlasma} shows how the integral $I(\nu)$ changes with $\Theta_{\rm pl}$ ($0 \le \Theta_{\rm pl} \le 10$) for fixed $\Theta_{\rm bm}$.
Three panels in Figure \ref{fig:ChangeThetaPlasma} correspond to different fixed values of $\Theta_{\rm bm}$ and the bottom-right sketch describes scattering geometry when $\Theta_{\rm bm} = 10$ corresponding to the bottom-left panel in Figure \ref{fig:ChangeThetaPlasma}, for example.
It is common for all the panels that `line a' is very close to `line b', i.e, we can approximate that the photon and plasma are completely aligned ($\Theta_{\rm pl} = 0$) even for $\Theta_{\rm pl} < 1$.
It is also common for all the panels that `line a' is larger than other lines for $\nu > \nu_0$ and $|I(\nu)|$ decreases in order from `line a' to `line e', i.e., $|I(\nu)|$ is large when the photons and the plasma are aligned at least the frequency range $\nu > \nu_0$.
The top-left panel ($\Theta_{\rm bm} = 0.1$) shows the case when the photon beam is considered as narrow (compared with $\gamma^{-1}$ cone associated with the plasma) and shows little frequency shift $D/D_1 \approx 1$ corresponding to $I_{\rm Narrow}$ and $I_{\rm Inclined}$ studied in Section \ref{sec:AnalyticEstimates}.
The bottom-left panel in Figure \ref{fig:ChangeThetaPlasma} is the case when the photon beam is considered as wide ($\Theta_{\rm bm} = 10$: the bottom-right sketch of Figure \ref{fig:ChangeThetaPlasma}).
In this case, the frequency shift effect is extreme and the absolute values $|I(\nu)|$ is almost unity at broad frequency range.

Figure \ref{fig:ChangeThetaPhoton} shows how the integral $I(\nu)$ changes with $\Theta_{\rm bm}$ ($0.1 \le \Theta_{\rm bm} \le 10$) for fixed $\Theta_{\rm pl}$.
Three panels in Figure \ref{fig:ChangeThetaPhoton} correspond to different fixed values of $\Theta_{\rm pl}$ and the bottom-right sketch describes scattering geometry when $\Theta_{\rm pl} = 1$ corresponding to the top-right panel in Figure \ref{fig:ChangeThetaPhoton}, for example.
Note that some lines are the same parameter set with Figure \ref{fig:ChangeThetaPlasma}.
It is common for all the panels that $|I(\nu)|$ decreases with the smaller values of $\Theta_{\rm bm}$.
`Line d' and `line e' on the top-left panel ($\Theta_{\rm pl} = 0$) and top-right panel ($\Theta_{\rm pl} = 1$) show $I_{\rm Wide}$ studied in Section \ref{sec:AnalyticEstimates}.

Lastly, we discuss the behaviors of `line d' and `line e' in the bottom-left panel in Figure \ref{fig:ChangeThetaPlasma} and of `line e' in the bottom-left panel in Figure \ref{fig:ChangeThetaPhoton}.
These lines satisfy $\Theta_{\rm bm} \ge \Theta_{\rm pl} > 1$ and shows two notable features.
One is the discontinuity point shifting toward $\nu < \nu_0$ (`feature one') and the other is $|I(\nu)|$ being significantly greater than unity at $\nu < \nu_0$ (`feature two').
We can discuss these features qualitatively.
To simplify explanation, we take $\Theta^2_{\rm bm} = \Theta^2_{\rm pl} \gg 1$ corresponding to `line e' in the bottom-left panel both in Figures \ref{fig:ChangeThetaPlasma} and \ref{fig:ChangeThetaPhoton} ($\Theta_{\rm bm } = \Theta_{\rm pl} = 10$).
For the `feature one', we obtain from Equation (\ref{eq:FrequencyShiftFactor}) that the frequency shift factor has a peak value $D / D_1 \sim \Theta^2_{\rm pl}$ at $\Theta_1 = \Theta_{\rm bm}$ and $\phi_1 = 0$, this value corresponds to the frequency which gives the peak of $|I(\nu)|$.
For the `feature two', we try to evaluate $|I(\nu \approx \Theta^{-2}_{\rm pl} \nu_0)|$.
For $\nu \approx \Theta^{-2}_{\rm pl} \nu_0$, we obtain $\nu_1 = (D/D_1) \nu \approx (\Theta^2_{\rm pl} / (1 + \Psi^2_1)) \nu \approx \nu_0 / (1 + \Psi^2_1) \le \nu_0$ so that we take $S(\nu_1) \sim p_1 + 2$ and $T_{\rm b}(\nu_1) \approx T_{\rm b}(\nu_0) (1 + \Psi^2_1)^{-p_1 - 1}$.
Assuming that $R$ is a constant of order unity, we obtain,
\begin{eqnarray}
	I(\nu \approx \Theta^{-2}_{\rm pl} \nu_0)
	& \approx &
	- \frac{3 R S}{16 \pi} \int^{2 \pi}_0 \int^{\theta_{\rm pl}}_0 d \phi_1 \theta^3_1 d \theta_1 
	\frac{4 \gamma^4}{(1 + \Psi_1^2)^{p_1 + 3}} \nonumber \\
	& \approx &
	- \frac{3 R S}{4 \pi} \int^{2 \pi}_0 \int^{\Theta_{\rm pl}}_0 d \phi_1 d \Theta_1 
	\frac{\Theta^3_1}{(1 + \Psi_1^2)^{p_1 + 3}}.
\end{eqnarray}
Although this integral cannot be performed analytically, we find that the integrand has a peak value $\Theta^3_{\rm pl}$ at $\phi_1 = 0$ and $\Theta_1 = \Theta_{\rm pl}$.
A crude estimate may be obtained by taking a peak value of the integrand $\Theta^3_{\rm pl}$ with $\int^{2 \pi}_0 d \phi \sim 2 \pi$ and $\int^{\Theta_{\rm pl}}_0 d \Theta_1 \sim \Theta_{\rm pl}$.
This must be overestimate and gives $3 R S \Theta^4_{\rm pl} / 2 \sim 10^4$ for $\Theta_{\rm pl} = 10$.
Although the value does not fit to the numerical calculation ($I(\Theta^{-2}_{\rm pl} \nu_0) \sim 10^2$ from Figures A1 and A2), we find the $I(\Theta^{-2} \nu_0)$ can be much greater than unity.

%

%

\end{document}